\documentclass[journal,10pt]{IEEEtran}

\usepackage{tikz,balance,algorithmic,algorithm}
\usetikzlibrary{shapes,decorations}
\usetikzlibrary{decorations.pathreplacing}
\newcommand{\ER}{Erd\H{o}s-R\'enyi}

\input{tcilatex}

\begin{document}

\title{Packet-Level Network Compression: Realization\\ and Scaling of the Network-Wide Benefits}

\author{
	Ahmad Beirami, \IEEEmembership{Student Member,~IEEE,}
	Mohsen Sardari, \IEEEmembership{Student Member,~IEEE, }\\
	Faramarz Fekri, \IEEEmembership{Senior Member,~IEEE}

\thanks{A. Beirami was with Georgia Institute of Technology. He is currently with the Department of Electrical and Computer Engineering, Duke University, Durham, NC 27708, USA. (email: ahmad.beirami@duke.edu).}

\thanks{M. Sardari and F. Fekri are with the School of Electrical and Computer Engineering, Georgia Institute of Technology, Atlanta, GA 30332-0250, USA. (email: \{mohsen.sardari, fekri\}@ece.gatech.edu).}

\thanks{
This material is based upon work supported by the National Science Foundation under Grant No. CNS-1017234.}
\thanks{
The material in this paper was partly presented at the IEEE Information Theory Workshop (ITW 2011)~\cite{ITW11} and IEEE International Conference on Computer Communications (INFOCOM 2012)~\cite{INFOCOM12}.}
}

\newcommand{\mc}{\mathcal}
\newcommand{\mb}{\mathbf}
\newcommand{\BH}{{\sf bit$\times$hop}}

\maketitle
\thispagestyle{empty}
\pagestyle{empty}
\begin{abstract}
The existence of considerable amount of redundancy in the Internet traffic at the packet level has stimulated the deployment of packet-level redundancy elimination techniques within the network by enabling network nodes to memorize data packets. Redundancy elimination results in traffic reduction which in turn improves the efficiency of network links. In this paper, the concept of network compression is introduced that aspires to exploit the statistical correlation beyond removing large duplicate strings from the flow to better suppress redundancy. 

In the first part of the paper, we introduce ``memory-assisted compression,'' which utilizes the memorized content within the network to learn the statistics of the information source generating the packets which can then be used toward reducing the length of codewords describing the packets emitted by the source. Using simulations on data gathered from real network traces, we show that memory-assisted compression can result in significant traffic reduction.

In the second part of the paper, we study the scaling of the average network-wide benefits of memory-assisted compression. We discuss routing and memory placement problems in network for the reduction of overall traffic. We derive a closed-form expression for the scaling of the gain in Erd\H{o}s-R\'enyi random network graphs, where obtain a threshold value for the number of memories deployed in a random graph beyond which network-wide benefits start to shine. Finally, the network-wide benefits are studied on Internet-like scale-free networks. We show that non-vanishing network compression gain is obtained even when only a tiny fraction of the total number of nodes in the network are memory-enabled. 
\end{abstract}



\vspace{-.1in}\section{Introduction}
\label{sec:intro}
Massive amount of data is daily produced and transmitted through various networks. A very high fraction of the cost of dealing with such massive data is associated with the transmission. Hence, any mechanism that can provide traffic reduction would result in huge cost benefits. 
Several studies have examined real-world network traffic and concluded the presence of considerable amounts of redundancy in the traffic data~\cite{Spring2000,anand_sigcomm_08,anand_sigmetrics_09,anand_sigcomm_09,SIVA-TECH-REPORT,shruti2012}. From these studies, \emph{redundancy elimination} has emerged as a powerful technique to improve the efficiency of data transfer in data networks.

Currently, the redundancy elimination techniques are mostly based on application-layer content caching~\cite{Rhea2003,compaction99}. However, several experiments confirm that the caching approaches, which take place at the
application layer, do not efficiently leverage the network traffic redundancy which exists mostly at the
packet level~\cite{anand_sigcomm_08,SIVA-TECH-REPORT}. Furthermore, caching approaches are incapable of suppressing redundancies that exist across multiple connections. They even lose opportunities for suppressing redundancy within one connection because of issues such as small content size. To address these issues, a few recent studies have considered the deployment of redundancy elimination techniques in the network layer (i.e., layer 3)~\cite{anand_sigcomm_08,SIVA-TECH-REPORT}, where the intermediate nodes in the network have been assumed to be capable of storing the previous communication in the network and also data processing. These works demonstrate that redundancy in the data is so high that even a simple de-duplication method (i.e., removing the repeated segments of the packets) can provide considerable bandwidth savings. Motivated by these benefits, in this work, packet-level redundancy elimination is investigated from an information-theoretic point of view.

Information theory has already established the fundamental limit in the compression of infinite-length sequences for the class of universal schemes~\cite{LZ77,CTW95, Baron_O_n}. Entropy is the fundamental limit (also called Shannon limit) of compression; sequences generated by a source cannot be compressed with a rate below entropy and uniquely decoded. A compression scheme is called universal if it does not require any prior knowledge about the source statistics. Hence, it is clear that from the practical point of view, the universal family is more interesting than the non-universal one. However, as shown in~\cite{Merhav_Feder_IT,ISIT11}, there is a significant penalty, i.e., gap from the asymptotic limit, when finite-length sequences are compressed under a universal scheme. More precisely, there is no hope of developing source coding algorithms that can compress short sequences to their entropy~\cite{ISIT11}. Also, in~\cite{ISIT11,ISIT12_gain}, we showed that the compression of small network packets requires more than 100\% compression overhead, beyond the Shannon limit. There is no way to get around this limit in the absence of side information.
 
 It is natural to ask whether it is possible to improve compression rates by taking advantage of side information about the source provided by the memorized sequences from previous traffic. In the finite block length regime of practical interest, it was demonstrated that by adding only 4MB of memory at the router it is possible to learn enough about a {stationary} source to overcome the fundamental limits of universal compression in the finite-length regime and approach the Shannon limit~\cite{ISIT12_gain,IT12_gain}. This is referred to as memory-assisted universal compression, where  a side information sequence of length $m$ (that can be thought of as the aggregate of the previous traffic stored in the network node) is obtained from the information-generating source. This length $m$ sequence that is stored both at the encoder (i.e., the source) and the decoder (i.e., the memory-enabled network node) is called ``memory'' and conveys important information about the source statistics. 

Even if the gain of memory-assisted compression is justified on a link, it remains an open problem how such a gain would scale if memory-assisted compression is adopted in the Internet on the router level. Several studies have inferred that the Internet (at the router level) can be very well modeled using a scale-free network~\cite{Albert1999,Faloutsos1999,Broder2000,Internet1,Internet2}. In particular, these studies provide evidence that although the Internet is growing dynamically, its properties can be well modeled through a stationary state described using the scale-free network model. Therefore, the end goal of this paper would be to understand the scaling of the network compression gain on Internet-like scale-free networks.

In the first part of this paper (through Section~\ref{sec:validation}), we discuss practical algorithms for implementation of {\em memory-assisted compression} and confirm the non-trivial gain of memory-assisted compression on data gathered from real network traces. This  is the achievable improvement in a link between the encoder and the decoder which both have a shared side information (or memory) of the previous communication and compare against the existing redundancy elimination techniques.
The memory-assisted compression is performed using both dictionary-based compression, e.g., gzip (also referred to as LZ77)~\cite{LZ77} and statistical compression, e.g.,  LPAQ~\cite{paq}, and context tree weighting (CTW)~\cite{CTW95}. These results provide a foundation for the network compression, which extends the idea of memory-assisted compression beyond a single link.

In the second part of this paper (from Section~\ref{sec:net-comp}), we extend our work to find achievable network-wide gain of memory-assisted compression (also referred to as network compression gain). In this context, we demonstrate that memory placement in the network poses some challenges to traditional shortest path routing algorithms, as the shortest path is not necessarily minimum cost route in networks with memory requiring modification in routing.
Further, we show that optimal memory placement in a network is non-trivial and vital to achieving network-wide benefits from memory deployment. 
To determine how the network compression gain scales with the number of memory-enabled nodes in the network, we theoretically quantify the scaling behavior of the benefits of memory-enabled nodes, Erd\H{o}s-R\'enyi (ER) random network graphs~\cite{erdos}. Considering ER random network graphs simplifies the problem significantly as the memory placement problem is trivial due to the symmetry and the analysis yields to a closed-form solution. The exact analysis reveals that there exists a threshold value for the number of memories deployed in a random graph below which the network-wide gain of memorization vanishes and above which it is fully accessible.
Finally, network compression gain is studied in Internet-like scale-free networks (which are modeled in this paper using random power-law graphs). Through analysis on random power-law graphs, it is demonstrated that non-vanishing network-wide gain of memorization is obtained even when the number of memory units is a tiny fraction of the total number of nodes in the network. 

Our contributions in this paper are summarized below.
\begin{itemize}

\item The concept of memory-assisted compression for redundancy elimination is introduced and its benefits are validated on real network data. In particular, memory-assisted compression gain is also defined which is the fundamental benefit that is obtained from memory in the network packet compression.

\item Network compression gain is defined and optimal routing strategy and memory placement algorithms in the presence of memory-enabled nodes are presented to maximize the network compression gain when the objective is the reduction of the aggregate traffic in the entire network.

\item The {\em average case} scaling of network compression gain is studied on Erd\H{o}s-R\'enyi random network graphs. It is shown that even with deployment of memory-enabled nodes that scale sublinearly with the total number of nodes in the network, non-negligible benefit from network compression is achievable on the average. 

\item The {\em average case} scaling of network compression gain is studied on scale-free networks modeled using Internet-like random power-law graphs, and it is demonstrated that significant network-wide benefits are obtained when only a tiny fraction of the network nodes are memory-enabled.

\end{itemize}

The rest of the paper is organized as follows.
In Section~\ref{sec:related-work}, the related work in network traffic reduction (redundancy elimination) is reviewed. In Section~\ref{sec:memory-assisted}, memory-assisted universal compression is introduced. In Section~\ref{sec:validation}, the benefits of memory-assisted compression for redundancy elimination are investigated through simulations on data gathered from real network traces. In Section~\ref{sec:net-comp}, the issues of extending memory-assisted compression to a network are described and the network compression gain is defined. 
In Section~\ref{sec:wired-routing}, the optimal routing and memory placement are investigated for maximizing the network-wide gain.
In Section~\ref{sec:wired-erdos}, network compression is studied for {\ER} random graphs.
In Section~\ref{sec:wired-powerlaw}, network compression is investigated for random power-law graphs.
Finally, the conclusion is given in Section~\ref{sec:conclusion}.

\vspace{-.1in}\section{Related Work}
\label{sec:related-work}
\subsection{Content-Centric Networking}
Recently, there has been a lot of attention regarding the efficient utilization of memory units inside network.
One related line of work is the  content-centric networking (CCN)~\cite{Named_Data_Networking,Jacobson2009}.
  In CCN, content is segmented into individually addressable pieces, and these individually addressable data segments are cached in the network. However, there are several fundamental differences between our work on network compression and the previous research on CCN.
  The first difference is that our approach deals with the data itself, as opposed to the content name.
As a simple example, two independent servers generating the same content but with different names would still be able to leverage the memory-assisted compression, but not the CCN. 
The second difference is that CCN has a fixed granularity of a packet, whereas one of the core features of compression algorithms is their flexibility to find redundancy in the data stream with arbitrary granularity. In fact, it is suggested that packet level caching, which most of the current techniques are approximately reduced to, offers negligible benefits for typical Internet traffic~\cite{SIVA-TECH-REPORT}, due to this predefined fixed granularity.

\subsection{Redundancy Elimination Using De-duplication}
\label{sec:dedup}
Another very related line of work considers the benefits of the deployment of packet-level redundancy elimination in the network~\cite{anand_sigcomm_08,anand_sigmetrics_09,anand_sigcomm_09,SIVA-TECH-REPORT,Siva_ACM,shruti2012} . The mechanism considered for redundancy elimination is mainly based on de-duplication. The de-duplication mechanism identifies the largest chunk of data that appears in memory and replaces it with a pointer~\cite{bently99}. 
It is common for network traffic to contain large repeated blocks, e.g., traffic from users that watch the same video. 


The core to de-duplication is an efficient value-based fingerprinting algorithm that is used to identify repeated chunks of data. In~\cite{karp-rabin87}, Karp and Rabin originally presented their pattern matching algorithm for string searching; the algorithm answers whether a particular pattern sequence exists in a packet of length $n$ by an efficient value-based fingerprinting~\cite{Spring2000,manber94}. 
The finger printing process facilitates the de-duplication by providing an easy to compute function that can quickly lead to identification of duplicates in the packets.

De-duplication works  well when the redundancy across the packets is in the form of large repeated chunks from a previously communicated packet. However, redundancy in data exists beyond simple repetitions in the form of statistical dependencies between symbols. This motivates our information-theoretic investigation of more efficient redundancy elimination algorithms based on data compression that target to leverage these statistical dependencies. 
We stress that the benefits of de-duplication and memory-assisted compression are complementary to each other as de-duplication provides a fast and efficient way of removing large repetitions whereas such repetitions would go mostly undetected using memory-assisted compression.  On the other hand, universal compression targets to leverage statistical dependencies between symbols in a sequence that would not be well detected using the de-duplication techniques. This observation is confirmed experimentally in Section~\ref{sec:ex-Shruti}.

\subsection{Compression (Source Coding)}
While relevant, the network compression problem is different from those addressed by distributed source coding techniques (i.e., the Slepian-Wolf problem) that target multiple correlated sources sending information to the same destination~\cite{slepian_wolf,Mina_TCOM}. In the Slepian-Wolf problem, the gains are achievable in the asymptotic regime. Further, the memorization of a sequence that is statistically independent of the sequence to be compressed can result in a gain in memory-assisted compression, whereas in the Slepian-Wolf problem, the gain is due to the bit-by-bit correlation between the two sequences.

Finally, as described in the introduction, fundamental limits of finite-length packet compression using memory-assisted compression were theoretically studied in~\cite{ISIT11,ISIT12_gain,IT12_gain} for a single source-destination link. 
This work extends the concept of memory-assisted compression to networks and aims at investigating the achievable network-wide compression benefits. Network compression requires new types of compression techniques that would take into account what is already memorized in the memory units and try to achieve the abovementioned fundamental limits laid down in~\cite{ISIT11,ISIT12_gain,IT12_gain}. 

\subsection{Network-Wide Issues of Redundancy Elimination}
In~\cite{anand_sigcomm_08, anand_sigcomm_09}, Anand {\em et al.} discussed network-wide issues of implementing redundancy elimination techniques, where they devise redundancy elimination-aware routing techniques for ISPs. This is done for traffic engineering objectives more advanced than the conventional minimum {\BH} routing. 
In contrast, we demonstrate that even addressing the minimum {\BH} routing is non-trivial for memory-assisted compression and leave the extension to more advanced traffic engineering as an open future research direction.
Since memory-assisted compression is complementary to the de-duplication based redundancy elimination, some of the solutions of~\cite{anand_sigcomm_08} for the latter probelm could be adapted to the network compression problem as well.
We stress that the goal of this paper is to investigate the average case scaling behavior of the network compression in an abstract sense as a function of the number of nodes in scale-free networks assuming that the practical implementation would be feasible. 

As discussed in~\cite{anand_sigcomm_09}, redundancy elimination in a network-wide setting results in non-trivial deployment challenges. First, since redundancy elimination requires the intermediate nodes to operate on the incoming packets, any solution must take into account the physical constraints and limitations of the available resources to carry out such operations. As such, in Section~\ref{sec:tradeoff}, we address the tradeoff between compression rate (performance) and compression complexity (required resources) of the developed memory-assisted compression. Another issue that needs to be addressed is that the architecture should be flexible so that network operators can choose and meet their overall network-wide goals using redundancy elimination. Although, in this paper, we only focus on overall traffic reduction as the simplest objective of network compression, more general objectives can be considered for network compression leading to different routing and memory deployment strategies. Lastly, the architecture should be designed in such a way that it can be adopted in an incremental fashion, i.e., you can equip network nodes with redundancy elimination capability one-by-one. We believe this latter issue can be addressed using the same techniques developed in~\cite{anand_sigcomm_09} for redundancy elimination based on de-duplication.



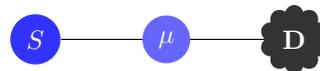
\begin{figure}[t]
\vspace{.1in}
\centering
\begin{tikzpicture}
\draw (0,0) node[anchor=east, circle, fill=blue!80, text=white]{$S$}
-- (1.4,0) 	node[circle, fill=blue!60, text=white]	(m){ $\mu$}
-- (3.1,0) 	node[cloud, fill=black!80, text=white]	(D){{$\mathbf{D}$}};
\end{tikzpicture}
\caption{The basic source $S$, memory $\mu$, and destination $\mb{D}$ configuration, where $\mathbf{D}$ represents a set of clients receiving packets from $S$.}
  \label{fig:S-D}
\end{figure}

\vspace{-.1in}\section{Memory-Assisted Universal Compression}
\label{sec:memory-assisted}

Consider an information source node $S$ which generates content to be delivered in the form of packets to a destination (client) node $D\in \mathbf{D}$ connected to $S$ through memory node $\mu$, as shown in Fig.~\ref{fig:S-D}.
Further, the client nodes in $\mathbf{D}$ request various sequences from the source over time.
Let $x^n = (x_1,...,x_n)$ be a sequence of length $n$, where each symbol $x_i$ is from the alphabet $\mc{A}$. For example, for an 8-bit alphabet that has 256 symbols, each $x_i$ is a byte. Note also that $x^n$ may be viewed as a packet at the network layer generated by source $S$.
Let $E [l_n(X^n)]$ denote the expected length resulting from the universal compression of $x^n$. 

Here, we consider two compression scenarios, as follows:
\begin{enumerate}
  \item Universal compression of an individual sequence with no memorization (Ucomp), in which a traditional universal compression is applied on the sequence $x^n$ without context memorization, and
  \item Memory-assisted universal compression with side information (UcompM), in which the encoder (e.g., server $S$ in Fig.~\ref{fig:S-D}) and the decoder (e.g., at the intermediate node $\mu$ in Fig.~\ref{fig:S-D}) both have access to a common side information (memory) $y^m$ from the same information generating source (to be explained), and they utilize memory for compression of the sequence $x^n$.
\end{enumerate}
In Ucomp, the intermediate nodes simply forward source packets to the client $D$ in sub-network $\mb{D}$. As such, compression takes place in the source and decompression is performed in the destination. Assuming a universal compression at the source, $E [l_n(X^n)]$ would be the length of the compressed sequence, which has to travel within the network from $S$ to the destination $D$ through the intermediate node $\mu$. Since every client in $\mb{D}$ requests a different sequence $x^n$ (but statistically dependent) over time, the source must encode each sequence $x^n$ independently and route through $\mu$. 
As such, clearly Ucomp does not utilize the correlations across different packets.
Now, consider the second scenario in which the intermediate node $\mu$, while serving as an intermediate node for different contents destined for different clients, memorize the contents and also constructs a model for the source $S$. As both source $S$ and the intermediate node $\mu$ are aware of the previous content sent to another client in $\mb{D}$, they can leverage this common knowledge for the better compression of the new packets sent over the $S-\mu$ portion of the path.

\begin{figure}[t]
\begin{center}
  \includegraphics[width=0.9\linewidth]{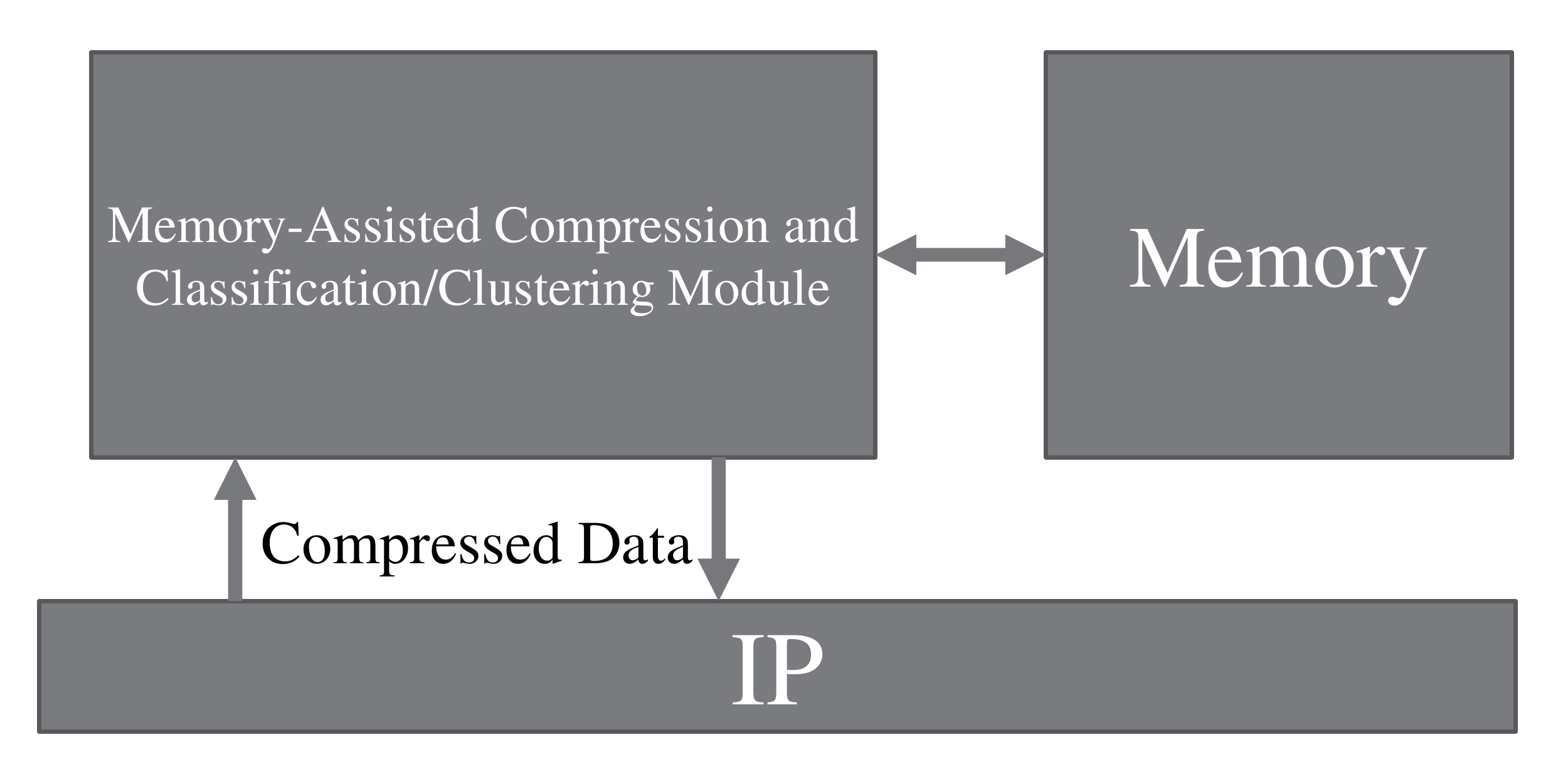}
  \caption{Network Compression architecture which includes the classification/clustering module.}
  \label{fig:netcomp-layer}
\end{center}
\vspace{-.1in}
\end{figure}

Here, perhaps, there is need for some clarifications. First, the memorization and learning from traffic takes place at the network layer because the routers (or the intermediate relays) are observing the packets at the network layer. Therefore, network compression should reside beneath the transport layer and above the network layer, at layer 3.5, as shown in Fig.~\ref{fig:netcomp-layer}.
Second, the intermediate node $\mu$ must decode and re-encode as the client at destination lacks memory, and hence, the client would not be able to decode a packet that is encoded using memory-assisted compression. This implies that if there are multiple routers or relay nodes on the path from the source to the destination, the last memory enabled router (i.e., the one that is closest to the client) must decode the packet using memory-assisted decoding and (possibly) re-encode the result using traditional universal compression before forwarding it to the client. Third, it is reasonable to assume that the client often lacks memory with the source. This is because the client is not connected to the source as often, and hence, even if it has obtained some packets from the source in the past, they may be outdated to carry information about source contents. Whereas, the routers are to observe the source packets much more often and hence have memorized and learned the source contents. Therefore, due to lack of memory at the client, the memory-assisted compression should not be applied end-to-end; from the source all the way to the client.

Specifically, assume that previous sequences (packets) $x^{m_1}, \ldots, x^{m_L}$ are sent from $S$ to clients $D_1, \ldots D_L$ in the sub-network $\mb{D}$ via $\mu$. Under UcompM, the node $\mu$ constructs a model for the source $S$ by observing the entire length $m = m_1 + \ldots +m_L$ sequence. Note that forming the source model by node $\mu$ is not a passive storage of the sequences $x^{m_1},\ldots,x^{m_L}$. This source model would be extracted differently for different universal compression schemes that will be used as the underlying memory-assisted compression algorithm. UcompM, which utilizes the memorized sequences of total length $m$, strictly outperforms Ucomp. This benefit, offered by memorization of the previous traffic as side information at node $\mu$, would provide savings on the amount of data transferred on the link $S-\mu$ without incurring any penalty except for some linear computation cost at node $\mu$. Please note that the memorization is used in both the encoder (the source) and the decoder (node $\mu$). Thus, source model is available at both $S$ and $\mu$.
From now on, by memory size we mean the total length $m$ of the observed sequences from the source at the memory unit.
We also stress that the network compression gain only applies after the initial memorization phase in which the memory-enabled routers populate their memory with packets from previous communication.

Let $E[l_{n| m}(X^n,Y^m)]$ be the expected code length for a sequence of length $n$ given a side information sequence $Y^m$ of length $m$ that is available to both the encoder and the decoder. The gain of memory-assisted compression $g\left(n,m\right)$ is defined as
\begin{equation}
g\left(n,m\right) \triangleq \frac{E[l_{n}(X^n)]}{E[l_{n|m}(X^n, Y^m)]}.
\label{eq:gain}
\end{equation}
In other words, $g(n,m)$ is the compression gain achieved by UcompM for the universal compression of the sequence $x^n$ over the compression performance that is achieved using the universal compression without memory (i.e., Ucomp) when the encoder and the decoder have a common side information sequence of length $m$.

The memory-assisted compression gain has been theoretically investigated in~\cite{ISIT12_gain, IT12_gain} where bounds on the achievable memory-assisted compression gain are provided for stationary parametric sources. It has been shown that with a memory size of 4MB from the same parametric source, it is possible to obtain more than two-fold gain in the compression rate of a new sequence. On the other hand, the purpose of this work is to validate such benefits on data gathered from real traffic traces using practical algorithms. Practical compression algorithms can be divided into two categories: statistical compression methods and the dictionary-based compression methods,  which are discussed in Sections~\ref{sec:statistical-comp} and~\ref{sec:dictionary-comp}, respectively.
In Section~\ref{sec:tradeoff}, we present the tradeoffs between the performance and complexity of memory-assisted compression algorithms.

%


%
\vspace{-.07in}\subsection{Statistical Compression Methods}
\label{sec:statistical-comp}
The essence of statistical compression methods is to find an estimate for the statistics of the source based on the currently observed sequence or an external auxiliary sequence. As such, the compression engine follows a two part design, a predictor followed by an arithmetic coder~\cite{Arithmetic_coding_introduction}. The predictor estimates the statistics of the source and a model is created using the previously seen symbols; based on this model predictions about the probability of the next symbol are issued. In short, the encoding of every new symbol entails:
\begin{enumerate}
  \item Estimating the likelihood of the symbol (e.g., byte) based on the model and context (previously seen symbols).
  \item Passing the estimated likelihood to the arithmetic encoder, which encodes the symbol.
  \item Updating the model using the new symbol.
\end{enumerate}
The decoding process is very similar to the encoding. Modern statistical compression algorithms such as Context Tree Weighting (CTW)~\cite{CTW95,CTW98}, Prediction by Partial Matching (PPM)~\cite{ppm-2002,ppm-Suzanne}, and PAQ~\cite{paq,compression-book} mix multiple simple models constructed sequentially to achieve better compression.

\subsubsection*{Context Tree Weighting (CTW)}
 A simple yet effective predictor can be constructed using tree models; Context Tree Weighting (CTW) algorithm is a well-known example of this approach~\cite{CTW95,CTW98}. CTW is used in part of the experiments in this paper.
In CTW, a tree of fixed depth $\delta$ is formed to represent the source model; the nodes on the tree correspond to estimates for the statistics of the source. Each bit is compressed according to the previous $\delta$ bits called context. Context bits determine a path in the tree that leads to one of the leaf nodes. The probability of the next bit is predicted by the information stored in the leaf node. The predicted probability is then sent to a binary arithmetic coder for compression. The tree nodes along the path are then updated using the next bit.

The generalization of the CTW encoding/decoding algorithm for the case of memory-assisted compression is immediate. As previously discussed, in memory-assisted compression, a sequence from the source is available to both the decoder (at $\mu$) and the encoder (at $S$). This sequence is the concatenation of all the packets sent from $S$ to $\mu$ in Fig.~\ref{fig:S-D}. Therefore, using this sequence, a context tree can be constructed that will be further updated in the compression process.
Note that the source and memory node should always keep the context tree synchronized with each other. In practical settings, a simple acknowledgment mechanism (such as the one in TCP/IP) suffices for the context synchronization.

\subsubsection*{Lite PAQ (LPAQ)}

LPAQ is a ``lite" version of PAQ, about 30 times faster than PAQ8~\cite{paq} at the cost of some compression (but similar to high-end PPM compressors~\cite{ppm-2002,ppm-Suzanne}).  The input sequence is processed sequentially and bit-wise. It follows the two part design discussed in Section~\ref{sec:statistical-comp}. The predictor in LPAQ employs seven models: $k$-gram Markov models of orders 1, 2, 3, 4, 6, and a ``match" model, which predicts the next bit in the last matching context.  The independent bit probability predictions of
the seven models are combined by a mixer, then arithmetic coded.
The $k$-gram Markov models consist of the last $k$ whole bytes plus any of the 0 to 7 previously coded bits of the current byte starting with the most significant bit.

PAQ mixer works with a binary alphabet and emits the probability of the next bit being 1. The estimates are geometrically weighted~\cite{mixing2012} and combined.  It can be verified that PAQ solves an instance of iterative gradient descent~\cite{mixing2012,nonlinear-book}. When the input sequence is stationary, the weights can be shown to converge to the true statistics of the source. 

\vspace{-.07in}\subsection{Dictionary-based Compression Methods}
\label{sec:dictionary-comp}
Unlike the statistical compression methods that rely on the estimation of the source statistical parameters, dictionary-based compression methods select sequences of symbols and encode each sequence using a dictionary of sequences that is generally constructed using the previously compressed symbols. The dictionary may be static or dynamic (adaptive). The former does not allow deletion of symbols from the dictionary, whereas the latter holds symbols previously found in the input stream, allowing for additions and deletions of symbols as new input is being read.

\begin{figure}
\begin{center}
 \includegraphics[width=.4\textwidth]{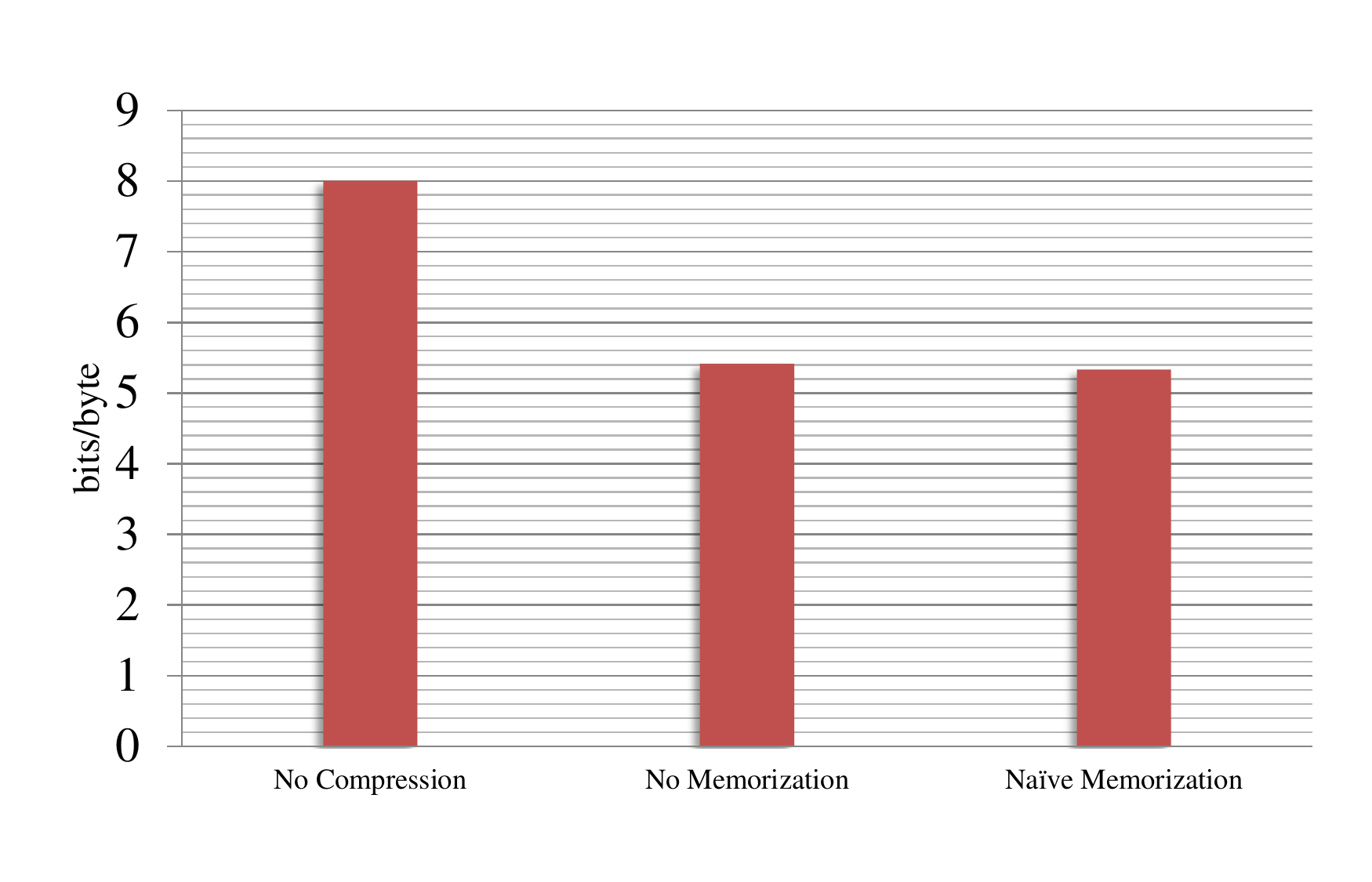}
\caption{The compression rate of a sample web trace (obtained from CNN web server) as a function of the sequence (i.e., packet) length, obtained using LZ77 and CTW compression algorithms.}
 \label{fig:motivation}
\end{center}
\end{figure}

\begin{figure*}[t]
\begin{center}
  \subfigure[Memory-assisted CTW]{
  \vspace{-.1in}
  \includegraphics[width=.37\textwidth]{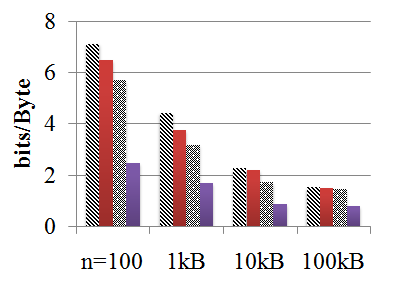}
  \vspace{-.1in}
  \label{fig:ctw}
  }
  \subfigure[Memory-assisted gzip (LZ77)]{
  \vspace{-.1in}
  \includegraphics[width=.37\textwidth]{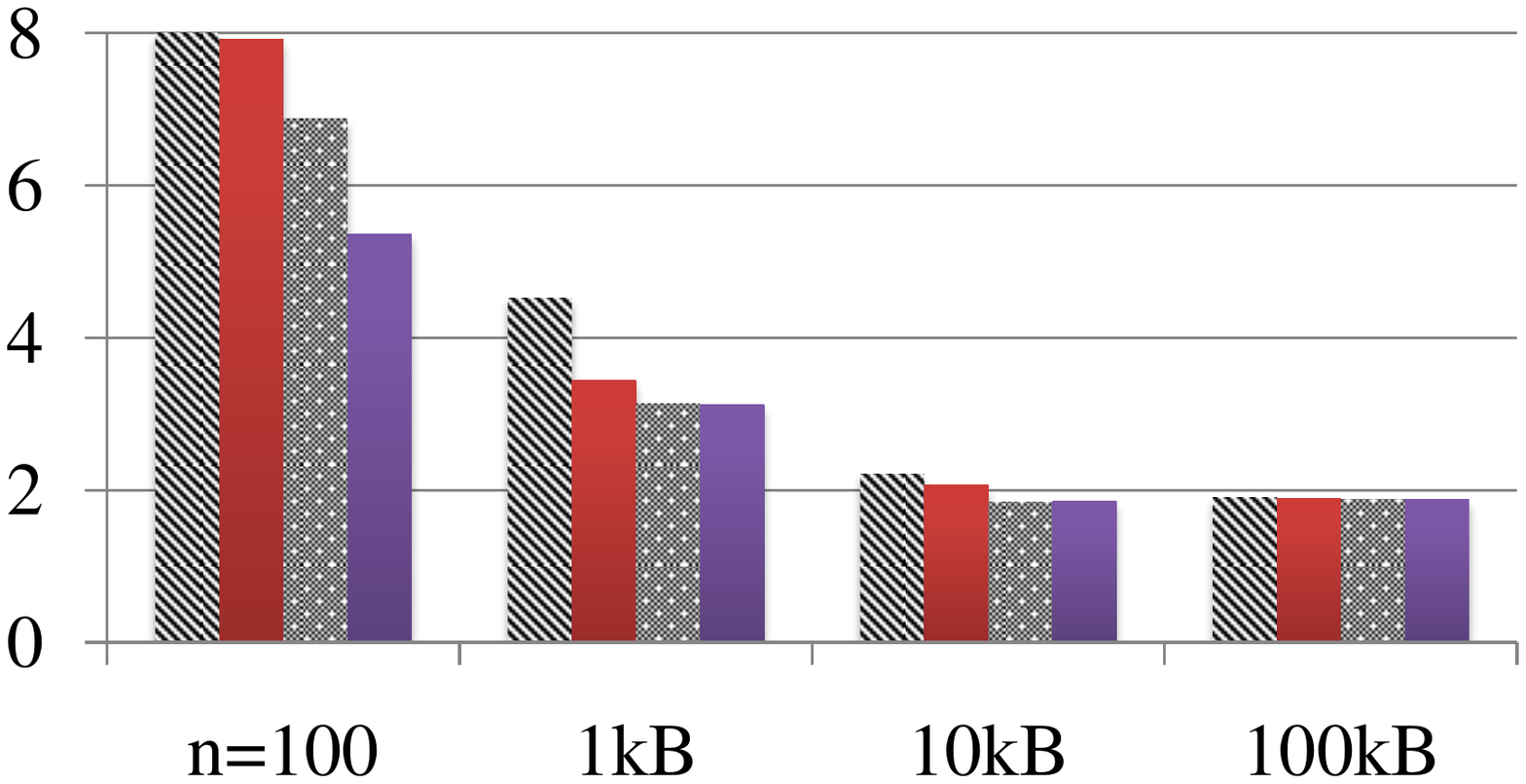}
  \vspace{-.1in}
  \label{fig:lz}
  }
   \subfigure{
  \includegraphics[width=.09\textwidth]{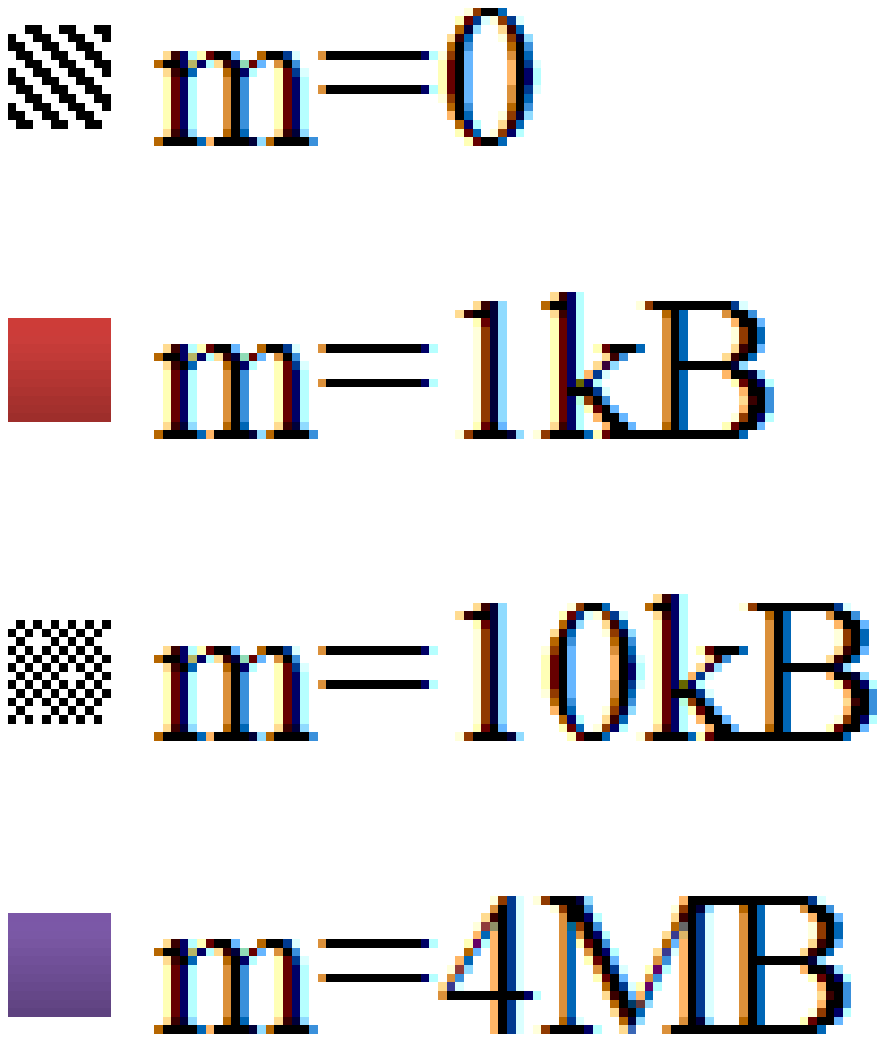}
  }
\end{center}
\vspace{-.1in}
\caption{The compression ratio (bits/Byte) achieved by memory-assisted universal compression algorithms.}
\vspace{-.1in}
\label{fig:memory-assisted-gains}
\end{figure*}

Dictionary-based compression algorithms have root in the seminal work of Lempel and Ziv~\cite{LZ77,LZ78}. This algorithm is based on a dynamically encoded dictionary that replaces a continuous stream of characters with codes. The symbols represented by the codes are stored in memory in a dictionary-style list. Dictionary-based algorithms are are widely used in practice and can be implemented efficiently and fast. However, the compression performance of these algorithms is considerably worse than a properly implemented statistical compression method.

\subsubsection*{gzip (LZ77)}
In this work, we use memory-assisted gzip, which is implemented based on the open-source DEFLATE algorithm. A sequence of length $m$ is assumed to be available at both the encoder and the decoder. The previously seen sequence is then used as the common dictionary. The new data to be compressed, is appended to the end of the dictionary at the source and fed to the gzip (LZ77) encoder. The output is sent to the decoder. Similarly, the decoder can reconstruct the intended stream by appending the transmitted symbols to the end of the dictionary and perform the gzip (LZ77) decoding algorithm.

\subsection{Compression Complexity}
\label{sec:tradeoff}
The speed and performance of different compression algorithms varies widely. The statistical compression algorithms are tailored to offer superior compression performance, however, the compression speed of this class of compression algorithms is considerably lower than dictionary-based compression algorithms. There are several high-speed dictionary based compression algorithms all of which can be considered variants of gzip, such as~\cite{gipfeli,Snappy,Fastlz,Quicklz}. These algorithms are the key part of some of the massive parallel computation systems, for example, Snappy~\cite{Snappy} is used in Google infrastructure. The main goal in the design of such high-speed algorithms has been to adapt LZ77 compression to achieve highest possible speed and through this process compression performance is traded for speed. As such, the compression performance of high-speed algorithms suffers, for example, compression of the first 1GB of the English Wikipedia using Snappy~\cite{Snappy}, and Gipfeli~\cite{gipfeli} has resulted in 530MB (in 2.8sec) and 410MB (in 4.3sec), respectively. However, the implementation of gzip that we experimented on would compress the same input to 320MB (in 41.7sec).\footnote{The execution time is measured on an Intel Xeon W3690 CPU. The execution time of the statistical compression methods is measured on a Intel core i5 processor using only one of the cores.} In contrast, PAQ8, CTW, and lite PAQ compress the 1GB English text input to 134MB (in $\approx$30 ksec), 211MB (in $\approx$13 ks), and 164MB (in $\approx$1 ksec), respectively. We note that improving compression speed and reducing the complexity of compression while maintaining acceptable compression rate is the subject of active research in the compression community (cf.~\cite{Baron-JSTSP} where the authors improve both compression performance and compression speed by using parallel compression). In conclusion, the high-speed compression algorithms are suitable where communication throughput is high and processing power is limited, e.g., 128MB/sec for Ethernet 1 Gigabit/sec connection. The high-performance statistical compression algorithms are more suitable for in a link where the communication speed is lower (6.5 MB/sec for 802.11g) and higher compression rates are desirable.

\section{Memory-Assisted Compression Gain on Real Network Traces}
\label{sec:validation}
\subsection{CNN Website Test Scenario}
\label{sec:ex-CNN}
In this section, we first demonstrate  the shortcomings of universal compression methods (without side information) for network packet compression.
The data used in this experiment is downloaded from CNN website (which is mostly text and script files).  To capture the packet, we used \emph{wget} and \emph{wireshark}~\cite{wireshark} open-source packet analyzer together and stored the IP packets.
We captured more than 18,000 data packets from the website. All packets have the same size of 1,434
bytes. In the first part of the experiment, we concatenated the packets to derive varying size super-packets and applied gzip (LZ77) and CTW on them.

As shown in Fig.~\ref{fig:motivation}, a modest compression performance can be achieved by compression of a packet when the super-packet length $n$ is small to moderate size. For example, for a data packet of length $n=1$kB, the compression rate is about $5$ bits per byte. Note that the uncompressed packet requires $8$ bits per byte for representation. Observe that as the packet length $n$ increases (here we have concatenated several packets payloads to achieve varying size packets), the compression performance improves. For very long sequences, the compression rate is about 0.5 bits per byte. In other words, comparing the compression performance between $n=1$kB and $n=16$MB, there is a penalty of factor $10$ on the compression performance (i.e., $5$ as opposed to $0.5$). Please note that the main reason this data set is compressible by more than 10 times is that it mostly consists of text files and scripts.

Next, we applied memory-assisted versions of LZ77 and CTW on the same data packets
as depicted in Fig.~\ref{fig:memory-assisted-gains}. To obtain the results in this plot, the first 4MB worth of packets from the data is used as memory. Then, 100 sequence are chosen from the rest of the data for compression and the average performance is reported in Fig.~\ref{fig:memory-assisted-gains}. As expected, the size of the compressed sequence decreases as memory size $m$ increases. For example, for a data sequence of length $n=100$B (which is obtained by manually extracting 100B from the payload of the packet), without memory, the compressed sequence has an average length of $\approx 87$B, while using a memory of size $m=4$MB, this data sequences can be compressed on average to 31B; almost 3 times smaller. This validates the theoretical predictions in~\cite{ISIT12_gain, IT12_gain} about the improvements achieved using memory-assisted compression.

\begin{figure}
\begin{center}
  \includegraphics[width=0.7\linewidth]{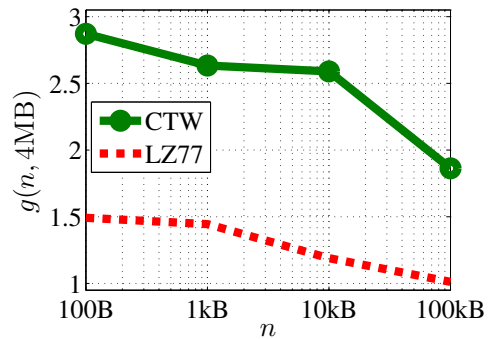}
\end{center}
\vspace{-.1in}
  \caption[The gain $g$ of memory-assisted compression over traditional compression (Ucomp), for memory size of 4MB for CTW and LZ compression algorithms.]{The gain $g$ of memory-assisted compression over traditional compression (Ucomp), for memory size of 4MB for CTW and LZ compression algorithms. This gain is achieved by utilizing memory on top of the performance of the conventional compression.}
  \label{fig:gain}
\vspace{-.1in}
\end{figure}

The actual gain of memory-assisted compression $g$ (defined in~\eqref{eq:gain}) for memory size 4MB is depicted in Fig.~\ref{fig:gain}. As can be inferred, the memory-assisted CTW (which is a statistical compression method) outperforms memory-assisted LZ77 (which is a dictionary-based method) in both the absolute size of the compressed output and also the gain of memory, i.e., the gain achieved on top of the gain of conventional compression, by utilizing memory. On the other hand, as discussed in Section~\ref{sec:tradeoff}, the suitable compression method has to be chosen based on the specific compression complexity requirements that need to be satisfied.


\subsection{Wireless Users Test Scenario}
\label{sec:ex-Shruti}
In this test scenario, we consider the data set used in~\cite{shruti2012}. The data set includes network traffic collected from 30 different mobile users consisted of smartphone users and laptop users. The laptop users relied only on WiFi connectivity for their network access. The smartphone users relied on both WiFi and 3G connectivity. The data collection spanned a period
of 3 months and yielded over 26 Gigabytes of unsecured down link data. Users accessed the Internet as per their normal behavior. More details about the acquisition process can be found in~\cite{shruti2012}.

\begin{figure*}
\begin{center}
  \includegraphics[width=\textwidth]{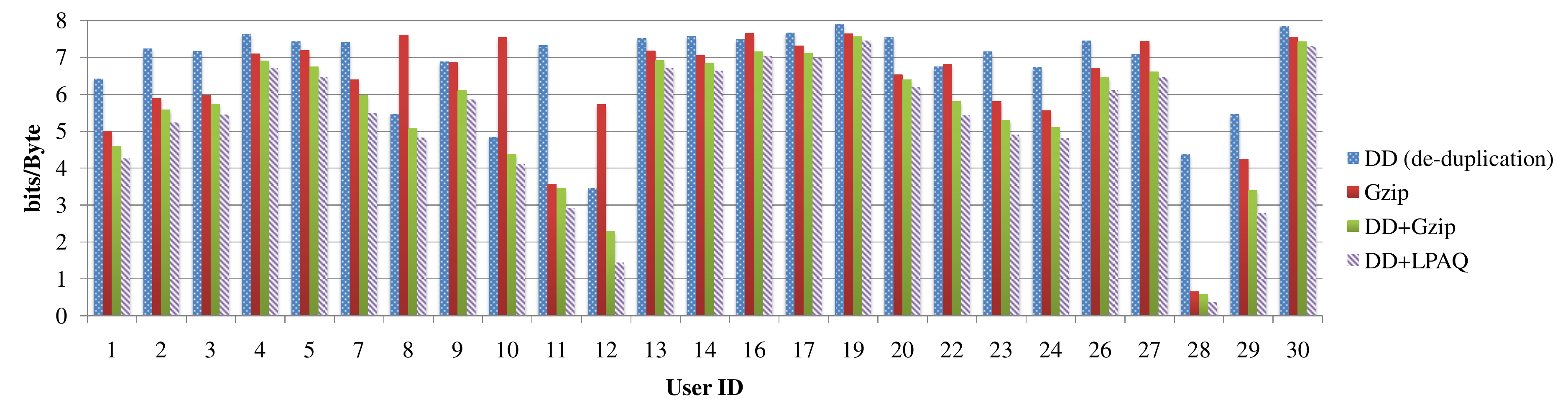}
  \caption{Performance of various compression algorithms on the data set used in the wireless users test scenario in Section~\ref{sec:ex-Shruti}.}
  \label{fig:dedup-all}
\end{center}
\end{figure*}

We also used the implementation of de-duplication from~\cite{shruti2012} for the sake of comparison with memory-assisted compression. The results are presented in Fig.~\ref{fig:dedup-all}, where the first bar shows the outcome of de-duplication, denoted by DD in short, and the second bar is the compression result of the trace data using gzip algorithm. 
The results are presented for each user in the data set. The packet size in each trace varies and is not fixed, however the packets are always less that 1,500 bytes. Each trace of each user is processed individually and the compression is performed on the entire trace data.  As expected, the performance of varies for different users.  In particular, in some cases, redundancy elimination is capable of reducing the input size by half whereas in some other cases little reduction can be observed.

Although gzip slightly outperforms de-duplication on most users, it is important to note that compression-based redundancy elimination in some cases performs worse than the de-duplication, e.g., on traces of Users 10 and 12. 
To understand this difference in behavior, we further analyzed the data for Users 11 and 12. For User 11, gzip outperforms de-duplication and for User 12, it is the other way around.
By analyzing the contents, we discovered that for User 11 long repeated sequences are scarce. Hence, the redundancy mostly exists in the form of statistical dependencies that can be captured using compression-based methods.
On the other hand, the data for User 12 abounds with long duplicates that are of tens of megabytes long. In such cases, de-duplication can more efficiently eliminate the existing redundancy. 
This confirms that de-duplication and compression target different types of redundancies in the data.

Next, we consider the interplay between de-duplication and memory-assisted compression
As described in Section~\ref{sec:dedup}, de-duplication can be used to identify and remove the long repeated chunks of data. Removing these repeated chunks before feeding the input sequence to the compression engine provides two major benefits for memory-assisted compression. First, the use of relatively fast de-duplication can speed up the high-performance but low-speed statistical data compressors. Second, the large repeated patterns would impact the predictor as it will try to learn and adapt to them while these patterns are not part of the existing statistical dependencies.  

The results of applying de-duplication before memory-assisted compression are presented in Fig.~\ref{fig:dedup-all}.
As can be seen, the combination consistently outperforms both de-duplication and memory-assisted compression on all traces. Further, Table~\ref{tab:dedup-summary} presents the average traffic reduction over all users. 
As can be seen, de-duplication achieves 15\% traffic reduction over these users compared with 22\% achieved by gzip memory-assisted compression. What is perhaps more interesting is the fact that when the two are applied in tandem, 30\% traffic reduction is achieved reaffirming that the two techniques target different types of redundancy in the data. 
Further, as expected the best performance is attributed to de-duplication in tandem with LPAQ memory-assisted compression achieving around 34\% traffic reduction on this data set.

\begin{table}
\centering
\caption[Summary of the compression rate (bits/byte) of various de-duplication and memory-assisted compression algorithms.]{Summary of the compression rate (bits/byte) of various de-duplication and memory-assisted compression algorithms.}
\begin{tabular}{|c|c|c|c|}
\hline
\textbf{De-duplication (DD)}	& \textbf{Gzip} & \textbf{DD+Gzip}	&	\textbf{DD+LPAQ} \\
\hline
6.80	&	6.24	&	5.61 	&	5.27	\\
\hline
\end{tabular}
\label{tab:dedup-summary}
\end{table}




\subsection{Memory Requirements}
To investigate the impact of the amount of physical memory on the compression performance, we derived the average compression performance of gzip, CTW, and LPAQ for different memory sizes. 
By increasing the memory size of gzip beyond 4MB (which is its standard memory size), compression performance improves and gets to close to the statistical compression methods with the cost of added complexity, which scales linearly with the memory size. On the other hand, the main purpose that gzip may be adopted is when fast compression is desired. For the statistical compression methods, we discovered that in both CTW and LPAQ, less than 2\% performance improvement is achieved when the memory size is increased from 4MB to 800MB (on the wireless users data set). Therefore, in this paper, in most of the experiments we chose to use a memory of size 4MB. 

We stress that it is reasonable to expect that a router that is sitting in the core of the network might see traffic from several different sources (users), which might be drastically different. Therefore, one might expect that the required memory size would then become much higher for such a router. For example, when a completely uncorrelated side information sequence is used for the compression of the current sequence, there is no hope in exploiting the side information for better compression. 
On the other hand, in~\cite{Allerton14}, we showed if the side information data is sufficiently correlated, high compression gains would be expected for a broad range of degrees of correlation. Further, in~\cite{WCSP14}, we mixed the data from the wireless users (to simulate what happens to a core router that sees all the data). We observed that using only 5 models, we can very well compress any sequence from all of these sources. Hence, we believe that although 4MB memory would probably be insufficient for a core router, the required memory size would not be unbounded either.

\section{Network-Wide Gain of Memory-Assisted Compression}
\label{sec:net-comp}
Thus far, we demonstrated that significant gain can be achieved using memory-assisted compression for redundancy elimination on the link level. The goal of the second part of this paper is to investigate how these benefits scale when memory-enabled nodes are deployed in a large-scale network. Specifically, the question that we are interested in answering is: ``Given the memory-assisted compression gain $g$, and a number of memory-enabled nodes capable of performing memory-assisted compression, and their placement, what is the achievable network-wide gain?''
Note that traffic traversing different paths in the network would see different gains at different time instances. On the other hand, we stress that our goal is present an average case study. Hence, in this section, we assume that each time a packet is compressed using memory-assisted compression, it experiences a gain $g$ (on the average). Although our assumptions do not provide much information about an individual traffic packet, we can draw conclusions about the {\em average} traffic reduction in the entire network and such conclusions become more and more relevant as the size of the underlying large-scale network grows larger. 



We represent a network by an undirected graph $G(V,E)$ where $V$ is the set of $N$ nodes (vertices) and $E=\{uv:u,v \in V\}$ is the set of edges connecting nodes $u$ and $v$. We consider a set of memory-enabled nodes $\boldsymbol{\mu} = \{ \mu_i\}_{i=1}^M$ chosen out of $N$ nodes where every memory node $\mu_i$ is capable of memorizing the communication passing through it. In this paper, as the first step, we assume that the total size of memorized sequences for each $\mu_i$ is assumed to be equal to $m$ and also assume that these nodes have similar resource constraints. 
Hence, we can assume that each memory unit will provide the same memory-assisted compression gain $g$ on the link from the origin node of the flow to itself.
The extension to meet resource limitations of individual nodes is left as an interesting future direction.

We focus on the expected performance of the network by averaging the gain over all scenarios where the source is chosen to be any of the nodes in the network equally at random.
In other words, we assume that the source is located in any node of the network uniformly at random, i.e., each node would be selected with probability $\frac{1}{N}$ and the destination is independently selected uniformly at random as well. 
In this paper, as the first step, we focus on minimizing the total cost of communication between the source and destinations in the network, measured by {\BH}. As will be shown, even this simplest objective raises non-trivial challenges. To meet more complex overall goals, one should refer to the techniques developed in~\cite{anand_sigcomm_09}.

Consider the outgoing traffic of the source node $S$ with the set of its destinations $\boldsymbol{D}= \{D_i\}_{i=1}^{N-1}$ each receiving different instances of the source sequence originated at $S$. Let $f_D$ be the unit flow from $S$ destined to $D \in \boldsymbol{D}$. The distance between any two nodes $u$ and $v$ is denoted by $d(u,v)$, which is measured as the minimum number of hops between the two nodes. As we will see later, introducing memories to the network will change the lowest cost paths, as there is a gain associated with the $S-\mu$ portion of the path. Therefore, we have to modify paths accounting for the memory-assisted compression gain. Accordingly, for each destination $D$, we define \emph{effective walk}, denoted by $W_D = \{S,u_1,\ldots,D\}$, which is the ordered set of nodes in the modified (lowest cost) walk between $S$ and $D$. Finding the shortest walk is the goal of routing problem with memories that minimizes the {\BH} cost.

We partition the set of destinations as $\boldsymbol{D} = \boldsymbol{D}_1 \cup \boldsymbol{D}_2$, where $\boldsymbol{D}_1=\{D_i:\exists {\mu}_{\scriptscriptstyle D_i} \in W_{D_i}\}$ is the set of destinations observing a memory in their effective walk, and 
\begin{equation}
\mu_{\scriptscriptstyle D_i} = \arg \min_{\mu \in \boldsymbol{\mu}} \{\frac{d(S,{\mu})}{g} + d({\mu}, {\scriptstyle D_i})\}.
\end{equation}
The total flow $\mathcal{F}_S$ of node $S$ is then defined as
\begin{equation}
\label{eq:totalflow}
\mathcal{F}_S \triangleq
\sum_{D_i \in \boldsymbol{D}_1}\left(\frac{f_{D_i}}{g}d(S,{\mu}_{\scriptscriptstyle {D_i}}) + f_{\scriptscriptstyle D_i}d({\mu}_{\scriptscriptstyle D_i},{\scriptstyle D_i} ) \right) \nonumber
\end{equation}
\begin{equation}
+ \sum_{D_j \in \boldsymbol{D}_2} f_{D_j} d(S,D_j).
\end{equation}
Using~\eqref{eq:totalflow}, we define $\hat{d}_D$, called the \emph{effective distance} from $S$ to $D$, as
\begin{equation}
\label{eq:effective_dist}
\hat{d}_D = \left\{
\begin{array}{ll}
\frac{d(S,{\mu}_{\scriptscriptstyle D})}{g} + d({\mu}_D, {\scriptstyle D})	& D\in  \boldsymbol{D}_1\\
d(S,D) 							& D\in  \boldsymbol{D}_2
\end{array}
\right.
.\end{equation}
In short, the effective distance is in the presence of gain $g$ obtained from memory-assisted compression. By definition, $\hat{d}_D \leq d(S,D)~ \forall D$.

In a general network topology, the network compression gain (denoted by $\mc{G}$) as a function of memory-assisted compression gain $g$ is defined as follows:
\begin{equation}
\label{eq:network-wide-gain}
\mathcal{G}(g)\triangleq\frac{\sum_{S\in V}\mathcal{F}^0_S}{\sum_{S\in V}\mathcal{F}_S}=\frac{\sum_{S\in V}\sum_{D\in \boldsymbol{D}} d(S,D)}{\sum_{S\in V}\sum_{D\in \boldsymbol{D}} \hat{d}_D}
,\end{equation}
where $\mathcal{F}^0_S$ is the total flow in the network by node $S$ without using memory units, i.e., $\mathcal{F}^0_S = \sum_{D\in \boldsymbol{D}} d(S,D)$. In other words, $\mathcal{G}$ is the gain observed in network achieved by memory-assisted scheme on top of what could be saved by universal compression (without memory) applied at the source and decoded at the destination. Alternatively, $\mc{G}(g)$ can be rewritten as
\begin{equation}
\frac{ \sum_{S\in V}\sum_{D\in \boldsymbol{D}} d(S,D)El_n(X^n)}{ \sum_{S\in V}\sum_{D\in \boldsymbol{D}} \left[d(S,{\mu}_{\scriptscriptstyle D})El_{n|m}(X^n) + d({\mu}_D, {\scriptstyle D})El_n(X^n)\right]}.
\end{equation}

\begin{figure}[t]
\centering
\begin{tikzpicture}
\draw (0,0) 	node[circle, fill=blue!90, text=white]	(S){$S$};
\draw (1,1) 	node[circle, fill=black!80, text=white]	(C1){{ }};
\draw (1,-1) 	node[circle, fill=black!80, text=white]	(C3){{$ $}};
\draw (3,1) 	node[circle, fill=black!80, text=white]	(C2){{$D_1$}};
\draw (3,-1) 	node[circle, fill=black!80, text=white]	(C4){{$ $}};
\draw (5,1) 	node[circle, fill=blue!70, text=white]	(m){{$\mu$}};
\draw (5,-1) 	node[circle, fill=black!80, text=white]	(C5){{$D_2$}};

\draw [very thick] (S)	--	(C1);
\draw [very thick] (S)	--	(C3);
\draw [very thick] (C1)	--	(C2);
\draw [very thick] (C2)	--	(m);
\draw [very thick] (C3)	--	(C4);
\draw [very thick] (C4)	--	(C5);
\draw [very thick] (C5)	--	(m);

\draw [->, red, thick, dashed] (0,0.55)	--	(1,1.55) 	-- 	(5.4,1.55) 	-- 	(5.4,0.45) -- (3, 0.45);
\draw [->, brown, thick, dashed] (0,.8)	-- 	(1, 1.8)  	--	(5.7,1.8)	--	(5.7,-1)	--	(5.5,-1);

\end{tikzpicture}
\caption{Example of routing in networks featuring memory: introducing memory-enabled nodes can lead to changes in the effective shortest paths (shown by dashed lines). Here, $g=3$.}
\label{fig:shortestpath}
\vspace{-.1in}
\end{figure}
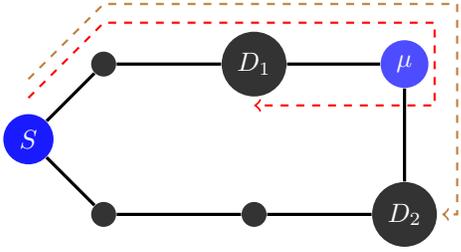

To demonstrate the challenges of the memory deployment problem and clarify the discussion, we consider one simple example network, which is presented in Fig.~\ref{fig:shortestpath}. Consider the destination node $D_1$, and let $g=3$. The effective walks from the source to destinations are obviously the shortest paths when there is no memorization (Ucomp coding strategy defined in Section~\ref{sec:memory-assisted}). As shown in the figure, when the node $\mu$ is memory-enabled, the effective path to $D_1$ changes from the shortest path. 
Prior to memory deployment, the shortest path to $D_1$ was two hops long, while enabling $\mu$ with memorization completely changes the effective distance to $D_1$ to be $\hat{d}_{D_1} = \frac{3}{3}+1 = 2$ as depicted in the figure. Note that in this example, we assumed no bandwidth constraint on the links. If $D_1$-$\mu$ link has a relatively small bandwidth, then the effective shortest path while minimizing the {\BH} might result in violating the bandwidth constraint on this link as the link would need to be used twice for passing one bit $D_1$. Taking such constraints into account would further complicate the problem and is not considered in this paper. The interested reader is referred to~\cite{anand_sigcomm_09} for further details on such issues.
Now, considering $D_2$, the cost of routing counter clockwise is 3, whereas it is $2 =\frac{3}{3}+1$ by passing through the node $\mu$ clockwise. This example shows that introducing memory-enabled nodes can result  in an effective shortest path that does not resemble the conventional shortest-path at all.

\vspace{-.1in}\section{Optimal Routing and Memory Placement Problem in Networks Compression}
\label{sec:wired-routing}
As demonstrated in the previous section, the deployment of memory-enabled nodes in the network gives rise to a number of questions and also brings some new challenges. 
We saw that the shortest path is not necessarily minimum cost route in networks with memory-enabled nodes, and hence, the well-known routing methods like Dijkstra's algorithm, in their original form are not optimal for networks with memory-enabled nodes. 

We aim at answering two fundamental (and related) questions regarding memory-assisted network compression.
In Section~\ref{sec:routing}, we derive the best strategy to route packets between the source and destination nodes given the network topology, the location of the memories, and the gain $g$ of memorization using a modification of Dijkstra's algorithm.
In Section~\ref{sec:placement}, we consider the problem of finding the best $M$ nodes (out of $N$) to maximize the benefits of memory deployment.


\vspace{-.07in}\subsection{Routing in Networks Featuring Memory}
\label{sec:routing}
 We consider an instance of network with a source node and fixed memory-enabled nodes and solve the routing problem in that instance of the network. Note that characterizing $\mathcal{G}$ involves computing both $\mathcal{F}^0$ and $\mathcal{F}$, which in turn requires finding the shortest paths and effective shortest paths between all pairs of nodes. The shortest path problem in a network without memory-enabled nodes is readily solved using Dijkstra's algorithm. 

Finding the shortest path using the Bellman-Ford algorithm relies on the so-called principle of optimality: if a shortest path from $u$ to $v$ passes through a node $w$, then the portions of the path from $u$ to$w$ and from $w$ to $v$ are also shortest paths. It is notable that in the networks with memory-enabled nodes the modified principle of optimality reads as: given a shortest path from $u$ to $v$, the portion of the path from $v$ to $w$ is still a shortest path whereas the portion from $u$ to $v$ is not necessarily a shortest path (This can also be seen in the example presented in Fig.~\ref{fig:shortestpath}).
Hence, we will use this modified principle of optimality to find the effective shortest path using the well-known Bellman-Ford algorithm which is obtained by repeatedly applying the principle of optimality. The Bellman-Ford algorithm is used in distance-vector routing protocols. The distributed version of the algorithm is used within an Autonomous System (AS), a collection of IP networks typically owned by an ISP. While Bellman-Ford algorithm solves the shortest path problem in networks with memory and the solution for routing within an AS, the more efficient Dijkstra's algorithm is used more widely in practice, most notably in \emph{IS-IS} and \emph{OSPF} (Open Shortest Path First) and hence we also visit the challenges of determining the effective shortest path in networks with memory-enabled nodes using the Dijkstra's algorithm.

The Dijkstra's algorithm solves the single-source shortest path for a network with positive edge costs, and hence, the {\BH} cost problem is a special case in which all the edge costs are equal to 1. 
Here, we present a modified version of Dijkstra's algorithm that determines the effective walk from all the nodes in a network to a destination $D$, in a network with a single memory-enabled node. Iterating over all nodes will provide the effective walk between every pair of nodes in the network. The extension to arbitrary number of memories is straightforward and skipped for brevity.
In a nutshell, to handle the memory node, we define a node-marking convention by defining a set $\mathcal{M}$ which contains the marked nodes. A node is marked if it is either a memory node, or it is a node through which a compressed flow is routed. The modified Dijkstra's algorithm starts with finding a node $\nu$ closest to node $D$. Then, it iteratively updates the effective distance of the nodes to $D$. The algorithm is summarized in Algorithm~\ref{algo:Dijkstra}. The notation $\text{cost}(vD)$, used in Algorithm~\ref{algo:Dijkstra}, is in fact the effective distance. At the beginning, the costs are initialized to $\text{cost}(vD)=\infty$ for nodes $v$ not directly connected to $D$, and then it is calculated for $v$ in every iteration. After finding the effective distance between every pair of vertices via the modified Dijkstra algorithm, we can calculate $\mathcal{F}$ and then determine the network compression gain $\mathcal{G}$.

\begin{algorithm}
\begin{algorithmic}
\STATE $\mathcal{M}=\mathbf{\mu}$
\WHILE {$V \neq \phi$}
	\STATE $\nu$ = the closest neighbor of $D$. 
\FOR {$\forall v \in V\setminus \{\nu,D\}$}
        \IF {$\nu \not\in \mathcal{M}$}
        	\STATE {$\text{cost}(vD) = \min\{\text{cost}(vD),\text{cost}(v\nu) + \text{cost}(\nu D)\}$}
        \ELSE
        	\STATE $\text{cost}(vD) = \min\{\text{cost}(vD),\frac{\text{cost}(v\nu)}{g} + \text{cost}(\nu D)\}$
        	\STATE $\mathcal{M}\gets\mathcal{M} \cup v$
        \ENDIF
\ENDFOR
\STATE $V\gets V \setminus \nu$
\ENDWHILE
\end{algorithmic}
\caption{Modified Dijkstra's Algorithm}
\label{algo:Dijkstra}
\end{algorithm}


\vspace{-.07in}\subsection{Memory Placement in a Network Graph}
\label{sec:placement}
The network compression gain depends on the number of memory-enabled nodes and also the locations they are deployed in the network. Since in practical scenarios only a select number of nodes have the storage and computational capability to function as a memory-enabled node, it is important to find the optimal location for such nodes.
Let the total number of memory units be $M$. The goal of the memory deployment is to find the best set of $M$ out of $N$ vertices in the network such that the network compression gain $\mathcal{G}(g)$ is maximized. It can be shown that this problem can be reduced to the well-known $k$-median problem, and hence, it is an NP-hard problem on a general graph. In other words, no tractable optimal placement strategy exists for a given general graph.

\begin{figure}[t]
\centering
\begin{tikzpicture}
\draw (0,0) 	node[circle, fill=blue!90, text=white]	(S){$S$}
-- (0.5,0) node[anchor= west, circle,draw, fill=black]{}
-- (1,0) node[anchor= west, circle,draw, fill=black]{}
-- (1.75,0) node[anchor= west,  fill=white]{$\ldots$}
-- (2.75,0) node[anchor= west, circle,draw,fill=black]{}
-- (3.5,0) node[anchor= west, circle,draw,fill=black]{}
-- (4.5,0) 	node[circle, fill=blue!70, text=white]	(m){ $\mu_i$}
-- (5.25,0) node[anchor= west, circle,draw,fill=black]{}
-- (6,0) node[anchor= west, circle,draw,fill=black]{}
-- (6.5,0) node[anchor= west,  fill=white]{$\ldots$}
-- (7.5,0) node[anchor= west, circle,draw,fill=black]{};

\draw [gray,decorate,decoration={brace,amplitude=5pt},xshift=0pt,yshift=-20pt]
   (4.5,0) -- (0,0)
   node [black,midway,below=4pt] { $t_i$};

\draw [gray,decorate,decoration={brace,amplitude=5pt},xshift=0pt,yshift=18pt]
   (2.8,0) -- (4.5,0)
   node [black,midway,above=4pt] { $\tau_i$};

\end{tikzpicture}
\caption{The placement of memory units on a line network: the source node $S$ is placed at one end of the line and the $i$-th memory is placed at $t_i$ from the source and $\tau_i$ is its left-coverage.}
\label{fig:line}
\vspace{-.1in}
\end{figure}
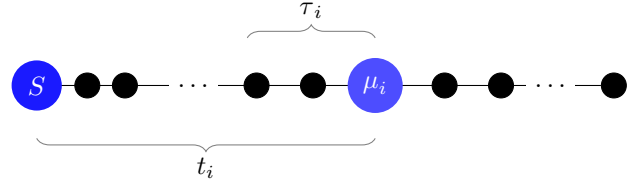

In this section, we demonstrate the challenges of the memory deployment problem by considering the class of line networks for which we can obtain closed-form solutions. Solving the  memory placement problem on this simple network topology will reveal why this problem is hard in general. 
Consider a line network with the source node $S$ placed at one end of the line and the destinations placed along the line as shown in Fig.~\ref{fig:line}. Therefore, we have a total number of $N$ nodes on the line and the total length of the line is $N$ hops.
As mentioned before, we assume traditional universal compression would give one unit of flow, to be sent to each destination. We consider the deployment of $M$ memory units on the line such that the memory $\mu_i$ is placed at hop-distance $t_i$ from the source. Without loss of generality, we also assume that $t_i < t_j$ for $i<j$, as shown in Fig.~\ref{fig:line}. We find $t_i$'s such that total flow $\mathcal{F}$ is minimized (or equivalently, $\mathcal{G}(g)$ is maximized). A related problem of finding ``en-route'' memory deployment on line networks is studied in~\cite{Krishnan2000}. En-route memories are those which are only located along routes from source to receivers. An en-route memory intercepts any request that passes through it along the regular routing path. The solution to the en-route memory placement problem as discussed in~\cite{Krishnan2000} is $t_i = \frac{i}{M} \quad \forall \mu_i.$

On the other hand, the memory deployment problem for network compression on a line network is more challenging. The difficulty arises from the fact that each memory-enabled node can serve some of the destination nodes that are located at a smaller hop-distance from the source than the memory node itself. As shown in Fig.~\ref{fig:line}, for a memory $\mu_i$ located at $t_i$, there is a left-coverage hop-length of $\tau_i$ towards the source to cover the destinations on the left side of the memory. The following lemma shows how $t$ and $\tau$ are affected when the memory-assisted compression gain $g$ chanes.

\begin{figure}[t]
\begin{center}
  \includegraphics[width=0.85\linewidth]{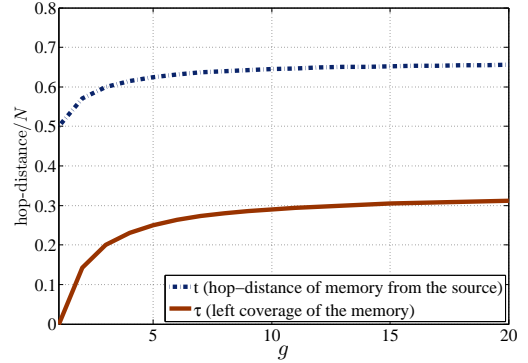}
\vspace{-.2in}
  \caption{Variations of $t$ and $\tau$ vs. $g$ for a line network.}
  \label{fig:tau-t}
\end{center}
\vspace{-.1in}
\end{figure}

\begin{lemma}
\label{lem:line-1}
For the case of $M=1$ and a line of hop-length $N$, the optimal memory location $t$ and coverage $\tau$ are given by
\begin{align}
\label{eq:line-1}
t 		&= \frac{2g}{3g+1}N + O(1), \\
\tau	&= \frac{g-1}{3g+1}N + O(1)
.\end{align}
\end{lemma}
\begin{IEEEproof}
The total flow can be written as
\begin{align}
\label{eq:line-proof}
\mathcal{F} = \int_0^{t-\tau} x \, \mathrm{d}x &+ \left ( \frac{t(1-t)}{g} + \int_0^{1-t} x \, \mathrm{d}x \right ) \nonumber \\
											   &+ \left (\frac{t\tau}{g} + \int_0^{\tau} x \, \mathrm{d}x \right )
\end{align}
The first term in~\eqref{eq:line-proof} is the flow to all points on the line not covered by memory. The second term is for the right coverage of memory and the third term accounts for the left coverage of memory ($\tau$). The result in~\eqref{eq:line-1} follows by taking the derivative of $\mathcal{F}$ and equating to zero, i.e., $\frac{\partial}{\partial t}\mathcal{F} = 0$ and $\frac{\partial}{\partial \tau}\mathcal{F} = 0$. 
\end{IEEEproof}
Figure~\ref{fig:tau-t} shows the plot of $t$ and $\tau$ versus $g$. as can be seen, as the gain $g$ increases, the optimal place for the memory is on $\frac{2}{3}N$ distance from the source and the left coverage approaches $\frac{1}{3}N$, whereas for $g=1$, the problem degenerates to that considered in~\cite{Krishnan2000}, i.e., the left-coverage is zero and the optimal place for memory is at $\frac{1}{2}N$.

\begin{lemma}
\label{lem:line-2}
The network-wide gain of placing a single memory ($M=1$) on a line network of bit-hop length $N$ as $N\to \infty$ converges to 
\begin{equation}
\mathcal{G}(g) = \frac{(3g+1)^2}{3g^2+10g+3}.
\end{equation}
Further, for $g\gg 1$ this converges to 
\begin{equation}
\lim_{g \to \infty} \mathcal{G}(g) = 3.
\end{equation}
\label{lem:2}
\end{lemma}
\begin{IEEEproof}
The proof is immediate from Lemma~\ref{lem:line-1} and the fact that for line $\mathcal{F}_0=1/2$.
\end{IEEEproof}
According to Lemma~\ref{lem:2}, if only one memory is deployed in the network, even if the memory-assisted compression gain is infinite (i.e., effectively no bits need to be sent from the source node $S$ to the memory node $\mu$), the network compression gain is finite. This reveals that to achieve proper network-wide scaling, the number of memory-enabled nodes would also need to scale. 

Following the results of deployment of a single memory on a line network, we can extend the result and solve for the general problem of deployment of $M$ memory-enabled nodes on a line network.

\begin{theorem}
\label{thm:thm-line}
Consider deployment of $M$ memory-enabled nodes on a line where memory, where $\mu_i$ is placed at $t_i$ and the left-coverage is denoted by $\tau_i$. Then,
\begin{equation}
\label{eq:thm1-1}
\left\{
\begin{array}{rcl}
t_i 	&= & 	\frac{i}{M} N + O(1)\\
\tau_i 	&= &	\frac{g-1}{2gM} N + O(1)
\end{array}
\right.
.\end{equation}
Furthermore as $N \to \infty$, we have
\begin{equation}\mathcal{G}(g) = \frac{2g^2M}{2g(M+1)+g^2+1},
\end{equation}
 and for $g \gg 1$ we have
\begin{equation}
\label{eq:line-gain}
\lim_{g \to \infty} \mathcal{G} = 2M
.\end{equation}
\end{theorem}
\begin{IEEEproof}
Similar to the proof of Lemma~\ref{lem:line-1}, we can write
\begin{align}
\mathcal{F} &= \sum_{m=1}^{M} \left [ \frac{t_i}{g} \tau_i + \int_0^{\tau_i} x \, \mathrm{d}x \right.\nonumber\\
 &+ \left. \frac{t_i}{g}(t_{m+1}-\tau_{m+1}-t_i)
 + \int_0^{t_{m+1}-\tau_{m+1}-t_i} x \, \mathrm{d}x \right ]
.\end{align}
Again, by taking the derivative of $\mathcal{F}$ with respect to $t_i$ and $\tau_i$ and solving the system of equations we arrive at
\begin{equation}
\label{eq:thm-line-proof}
\left\{
\begin{array}{rcl}
t_i 	&= &\frac{t_{i+1}+t_{i-1}}{2}  \\
\tau_i 	&=	&	\frac{g-1}{2g}\left (t_i-t_{i-1} \right )
\end{array}
\right.
,\end{equation}
where~\eqref{eq:thm-line-proof} results in a tridiagonal matrix which in turn results in~\eqref{eq:thm1-1} for large $M$. Further,~\eqref{eq:line-gain} follows from~\eqref{eq:thm1-1} and $\mathcal{F}_0=1/2$ for line networks.
\end{IEEEproof}

Thus far, we showed that the memory deployment problem is non-trivial even on a line network and the optimal solution is indeed not very intuitive at the first glance. Furthermore, these problems would need to be modified when other constraints such as link bandwidths are present~\cite{anand_sigcomm_09}.

\vspace{-.1in}\section{Network Compression in Erd\H{o}s-R\'enyi Random Network Graphs}
\label{sec:wired-erdos}
In this section, we would like to analyze the minimum number of memory-enabled nodes required for the network-wide benefits of memory-assisted compression to be achievable in a large-scale network. 
In one extreme, if all of the network nodes are memory-enabled, almost automatically, the link gain would translate to the network-wide gain. On the other hand, if only a constant number of nodes participate, the performance improvement would not scale properly (as observed also for a line network in Section~\ref{sec:placement}). This section aims at addressing what happens in between these extremes. 
To do so, we use Erd\H{o}s-R\'enyi (ER) random network graphs. The main reason is that the symmetry in these network graphs yield to analytic closed-form solution that can be insightful when considering general network graphs. 

\vspace{-.07in}\subsection{Background on ER Random Graphs}

Before we can state our main results, we need to cover the background on ER random graphs.
\begin{definition}
An ER random graph $G(N,p)$ is an undirected, unweighted graph on $N$ vertices where any two vertices are connected with an edge with probability $p$.
\end{definition}
\begin{definition}
Let $u,v \in G$ be any two vertices. The diameter of a connected graph is defined as $\max_{u,v} d(u,v)$. Further, the average distance of a connected graph is defined as $\mathbf{E}[d(u,v)]$.
\end{definition}

The following properties hold for ER random graphs~\cite{chung-book}:
\begin{enumerate}
  \item  $G(N,p)$ contains an average of ${N\choose 2} p$ edges.
  \item  If $p<\frac{(1-\epsilon)\log N}{N}$, then $G(N,p)$ almost surely (a.s.) has isolated vertices and thus disconnected.
  \item  If $p=\frac{c\log N}{N}$ for some constant $c>1$, then $G\left(N,p\right)$ is a.s. connected and every vertex asymptotically has degree $c\log N$~\cite{alon2008}.
  \item The diameter of $G(N,p)$ is almost surely $\frac{\log N}{\log Np}$.
  \item The average distance in $G(N,p)$, denoted by $\bar{d}$, is
  \begin{equation}
\label{eq:avg_dist}
\bar{d} = (1+o(1))\frac{\log N}{\log Np},
\end{equation}
provided that $\frac{\log N}{\log Np} \to \infty$ as $N \rightarrow \infty$ (this condition is satisfied in the connected regime).%
\end{enumerate}

\vspace{-.07in}\subsection{Main Result}
To characterize the network compression gain, we consider connected $G(N,p),~p=\frac{c\log N}{N}$, with a single source node $S$ and all other nodes as destinations (uniformly at random). Since the expected degree of all nodes in ER graph is equal and every vertex is chosen as a destination with equal probability, the optimal memory selection problem in this case disappears, and we select memories $\{\mu_i\}_{i=1}^M$ uniformly at random. Theorem~\ref{thm:scaling} provides the scaling of $\mathcal{G}(g)$ with respect to $M$.

\begin{theorem}
\label{thm:scaling}
Suppose $M$ memory-enabled nodes are deployed on an ER random graph and let $\epsilon >0$ be a positive real number. Then,
\begin{itemize}
  \item  [(a)] If $M = O\left(N^{\frac{1}{g}-\epsilon}\right)$, then $\mathcal{G}(g)\sim 1$. \footnote{Throughout this work, we have used the following asymptotic notations:
\begin{itemize}
  \item $f(x) = o(g(x))$ iff $|f(x)| \leq |g(x)| \epsilon,~\forall \epsilon$,
  \item $f(x) = O(g(x))$ iff $|f(x)| \leq |g(x)| k,~\exists k$,
  \item $f(x) = \Omega(g(x))$ iff $|f(x)| \geq |g(x)| k,~\exists k$, and
  \item $f(x) \sim g(x)$ iff $f(x)/g(x) \rightarrow 1$.
\end{itemize}}

  \item  [(b)] If $M = \Omega\left(N^{\frac{1}{g}+\epsilon}\right)$, then $\mathcal{G}(g) \sim  \frac{g}{1-g\log_N(\frac{M}{N})}$.
\end{itemize}
\end{theorem}
\begin{IEEEproof}[Sketch of the proof]
We first find an upper bound on the number of destinations benefit form each memory. This upper bound is sufficient to derive part (a) of the theorem. For the second part, we find a lower bound on the number of benefiting destinations. 
\end{IEEEproof}

To characterize $\mathcal{G}(g)$, we first need to find $\mathcal{F}_0$. The average distance from the source to a node is $\bar{d}$. Thus, $\mathcal{F}_0 = N\bar{d}$. For large $N$, \eqref{eq:avg_dist} results in
\begin{equation}
\mathcal{F}_0 \sim  \frac{N\log N}{\log \log N}.
\end{equation}

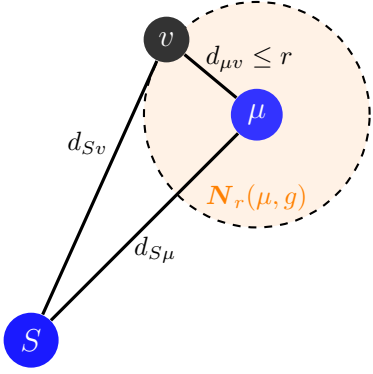
\begin{figure}[t]
\centering
\begin{tikzpicture}
\draw (0,0) 	node[circle, fill=blue!90, text=white]	(S){\large{$S$}};
\draw [thick, dashed, fill=orange!10, text=orange] (3,3) circle (1.5) (3,1.9) node {$\boldsymbol{N}_r(\mu,g)$};
\draw (3,3) 	node[circle, fill=blue!80, text=white]	(m){\large{$\mu$}};

\draw (1.8,4) 	node[circle, fill=black!80, text=white]	(v){\large{$v$}};

\draw [very thick] (S)	--	(m) (1.65,1.2) node {$d_{S\mu}$};
\draw [very thick] (m)	--	(v) (2.9,3.75) node {$d_{\mu v}\leq r$} ;
\draw [very thick] (S)	--	(v) (.75,2.62) node {$d_{Sv}$} ;

\end{tikzpicture}
\vspace{-.1in}
\caption{Illustration of Memory Neighborhood.}
\label{fig:neighbor}
\vspace{-.1in}
\end{figure}

Next, we need to find $\mathcal{F}$. For every memory $\mu$ we consider a neighbourhood $\boldsymbol{N}_r(\mu)$ as shown in Fig.~\ref{fig:neighbor}. This neighborhood consist of all vertices $v$ within distance $r$ from $\mu$. We choose $r$ such that, almost surely, all nodes in $\boldsymbol{N}_r(\mu)$ would benefit from the memory node $\mu$. Clearly, if $\frac{d(S,\mu)}{g} + r = d(S,v)$, the benefit provide by the memory for node $v$ vanishes and only nodes at distances less than $r$ benefit from the memory $\mu$. Given $g$, we denote this set of nodes benefiting from $\mu$ by $\boldsymbol{N}_r(\mu,g)$.
\begin{equation}
\label{eq:neibor}
\boldsymbol{N}_r(\mu,g) = \left\{v:\frac{d(S,\mu)}{g} + d(\mu, v) \leq d(S,v)\right\}
.\end{equation}
Since memory nodes are uniformly placed, the average value of $d(S,\mu)$ in $\hat{d}_v$ is equal to $\bar{d}$. Similarly, the average of $d(S,v)$ is also $\bar{d}$. Hence, solving for $r$ in~\eqref{eq:neibor} and then using the result on the average distance in~\eqref{eq:avg_dist}, we conclude
\begin{equation}
\label{eq:neigh-radius}
r \stackrel{a.s.}{=} (1-{1}/{g})\left(\frac{\log N}{\log \log N}\right) .
\end{equation}

The following lemma, by Chung and Lu~\cite{chung-book}, gives an upper bound on the total number of vertices in the neighborhood $\left |\boldsymbol{N}_r(\mu_i,g)\right |$, where $|\cdot|$ is the set size operator.
\begin{lemma}[\cite{chung-book}]
\label{lem:neighbor}
Assume a connected random graph. Then, for any $\epsilon>0$, with probability at least $1-\frac{1}{(\log N)^2}$, we have
$ \left |\boldsymbol{N}_r(\mu_i,g)\right | \leq (1+2\epsilon)(Np)^r$, for $1\leq r \leq \log N$.
\end{lemma}

Using Lemma~\ref{lem:neighbor} and~\eqref{eq:neigh-radius}, we deduce that
\begin{align}
\label{eq:gain-vertex}
|\boldsymbol{N}_r(\mu_i,g)| \stackrel{a.s.}{\leq} & (1+2\epsilon)(\log N)^{(1-\frac{1}{g})\left(\frac{\log N}{\log \log N}\right)} \\
= \hspace{0.03in}&(1+2\epsilon)N^{1-1/g}
.\end{align}
Therefore, the total number of nodes that gain from the memory-enabled nodes is upper-bounded by
\begin{equation}
\sum_{i=1}^M |\boldsymbol{N}_r(\mu_i,g)| \leq M(1+2\epsilon)N^{1-1/g}
.
\end{equation}
From~\eqref{eq:gain-vertex}, it is clear that the network compression gain vanishes if $M$ is too small. The value $N^{1/g}$ is the threshold value for the network-wide gain. More accurately, if $M = O\left(N^{\frac{1}{g}-\epsilon}\right)$, then the memory-assisted compression gain would not result in any network-wide improvement. This is in contrast to the line network where with a single memory, we would obtain network-wide improvements.

\vspace{-.07in}\subsection{Proof of the Main Result}
\begin{IEEEproof}[Proof of Theorem~\ref{thm:scaling}(a)]
For all the nodes in $\boldsymbol{N}_r(\mu_i,g)$, we have a flow gain of $g$. Let $M = N^{\frac{1}{g}-\epsilon}$, then we have
\begin{align}
	\label{eq:double_counting}
\mathcal{G}(g)
\leq \hspace{0.03in}& 		\frac{N\bar{d}}{\frac{\bar{d}}{g}M|\boldsymbol{N}_r(\mu,g)|+\bar{d}(N-M|\boldsymbol{N}_r(\mu,g)|)}\\
	\label{eq:insert-neighbor-size}
\stackrel{a.s.}{\leq}& 		\frac{N}{N-(1-1/g)M  N^{(1-\frac{1}{g})}} \\
= \hspace{0.03in}&			\frac{N}{N-(1-1/g)N^{1-\epsilon}} \\
\sim \hspace{0.03in}&		1 
,\end{align}
where inequality in~\eqref{eq:double_counting} follows from the double counting  of the destination nodes that may reside in more than one neighborhood. Also,~\eqref{eq:insert-neighbor-size} follows from replacing~\eqref{eq:gain-vertex} in~\eqref{eq:double_counting}.
\end{IEEEproof}

Since we need more than $ n^{\frac{1}{g}}$ memory units to have a network-wide gain, the next question is as to how $\mathcal{G}(g)$ scales when the number of memory units exceeds $ n^{\frac{1}{g}}$. To answer this question, we need to establish a lower-bound on the neighborhood size and the number of nodes benefiting from memory. Further, we have to account for the possible double counting of the intersection between the memory neighborhoods. We invoke the following concentration inequality from~\cite{chung-book} to establish the desired bound.
\begin{proposition}[\cite{chung-book}]
\label{thm:concetration}
If $X_1,X_2,\ldots,X_n$ are non-negative independent random variables, then the sum $X=\sum_{i=1}^n  X_i$ holds the bound
\begin{equation}
\boldsymbol{P}[X\leq \mathbf{E}[X]-\lambda]\leq \exp\left({-\frac{\lambda^2}{2\sum\mathbf{E}[X_i^2]}}\right)
.\end{equation}
\end{proposition}
This inequality will be helpful to show that the quantities of interest concentrate around their expected values.

The following lemma provides a lower-bound on the neighborhood size $\left |\boldsymbol{N}_r(\mu,g)\right |$ and the lower-bound on $\mathcal{G}(g)$, as we show, is immediate.
\begin{lemma}
\label{lem:lowerbound-neighbor}
Consider a set of vertices $V$ of $G(N,p)$ such that $\frac{|V|}{N}=o(1)$. For $0<\epsilon<1$, with probability at least $1-e^{-{Np|V|}{\epsilon^2}/2}$, we have
\begin{equation}
\label{eq:neighbor-lowerbound}
\left |\boldsymbol{N}_r(\mu,g)\right | \geq (1-\epsilon)(Np)^r.
\end{equation}
\end{lemma}
\begin{IEEEproof}
The vertex boundary of $V$, denoted by $\Gamma(V)$, consists of all vertices in $G$ adjacent to some vertex in $V$.
\begin{equation}
\Gamma(V)=\left \{u:u\not \in V, \text{ and }u \text{ is adjacent to }v\in V\right\}
.\end{equation}
Let $X_u$ be the indicator random variable that a vertex $u$ is in $\Gamma(V)$, i.e., $\boldsymbol{P}[X_u=1] = \boldsymbol{P}[u\in\Gamma(V)]$. Then,
\begin{align}
\mathbf{E}\left[|\Gamma(V)|\right] &= \sum_{u\not\in V} \mathbf{E}[X_u]\\
& = \sum_{u\not\in V}\boldsymbol{P}[u\in\Gamma(V)]\\
&=	\sum_{u\not\in V}\left(1-(1-p)^{|V|}\right)\\
\label{eq:expectedboundary-1}
&\geq p|V|(N-|V|)\\
& = (1+o(1))Np|V|
\end{align}
where the inequality in~\eqref{eq:expectedboundary-1} follows from
\begin{align}
\boldsymbol{P}[u\in\Gamma(V)]&	=	1-(1-p)^{|V|}\\
& \geq 1-e^{-p|V|} \\
&\sim p|V|
,\end{align}
and the second part holds because $\frac{|V|}{N}=o(1)$. Since, $X_u$'s are non-negative independent random variables, by applying Proposition~\ref{thm:concetration} with $\lambda = \sqrt{\alpha\mathbf{E}[|\Gamma(V)|]}$, with probability at least $1-e^{-\alpha/2}$ we have
\begin{align}
\label{eq:gamma-size}
|\Gamma(V)| &\geq \mathbf{E}\left[|\Gamma(V)|\right] - \sqrt{\alpha\mathbf{E}[|\Gamma(V)|]}\\
&\geq (1-\epsilon)Np|V|
.\end{align}
By picking a single vertex and applying~\eqref{eq:gamma-size} inductively $r$ times, and then adding up the number of adjacent nodes, we obtain~\eqref{eq:neighbor-lowerbound}.
\end{IEEEproof}

Now that we have a lower-bound on the number of nodes benefiting from each memory, we show that by increasing the number of memories beyond $M=N^{\frac{1}{g}}$, memories cover all the nodes in the graph effectively and hence all the nodes would gain from the memory placement.

In order to limit the intersection between the neighborhoods, we reduce $r$ to $r_\delta$ as below:
\begin{equation}
\label{eq:r-delta}
r_\delta = (1-{1}/{g}-\delta)\left(\frac{\log N}{\log \log N}\right) .
\end{equation}
With this choice of $r_\delta$, invoking Lemmas~\ref{lem:neighbor} and~\ref{lem:lowerbound-neighbor}, we deduce that the probability that a random node $u\in G$ belongs to the neighborhood $\boldsymbol{N}_{r_\delta}(\mu_i,g)$ of the memory $\mu_i$ is $N^{-1/g-\delta}$. Hence, the expected number of the covered nodes is
\begin{align}
\mathbf{E}\left[\left|\bigcup_{i=1}^M \boldsymbol{N}_{r_\delta}(\mu_i,g)\right|\right]	&= \sum_{u\in G}\boldsymbol{P}\left[u \in \cup_{i=1}^M \boldsymbol{N}_{r_\delta}(\mu_i,g)\right]\\
&=\sum_{u\in G}\left(1-(1-N^{-1/g-\delta})^M\right) \\
&\sim N\left(MN^{-1/g-\delta}\right)\\
&\sim N\label{eq:neighbor-delta}
,\end{align}
where~\eqref{eq:neighbor-delta} holds by choosing $M=N^{1/g+\delta}$.

To show that the number of covered nodes is concentrated around its mean, we use Proposition~\ref{thm:concetration} again with $\lambda = \sqrt{\alpha\mathbf{E}\left[\left|\cup \boldsymbol{N}_{r_\delta}(\mu_i,g)\right|\right]}$. Then, with probability at least $1-e^{-\alpha/2}$ we have
\begin{align}
\left|\bigcup_{i=1}^M \boldsymbol{N}_{r_\delta}(\mu_i,g)\right|	&\geq \mathbf{E}\left[\left|\cup \boldsymbol{N}_{r_\delta}(\mu_i,g)\right|\right]- \lambda\\
&\geq	(1-o(1))N
.\end{align}
Hence, the memory-enabled nodes cover, almost surely, all of the nodes.

Since all nodes are covered with high probability, we can associate each node with a neighborhood $|\boldsymbol{N}_{r_\delta}(\mu_i,g)|$, for which nodes' distances in the neighborhood from memory are $(1-o(1))r_\delta$.
\begin{IEEEproof}[Proof of Theorem~\ref{thm:scaling}(b)]
 By~\eqref{eq:neighbor-delta}, we can bound the network-wide gain of the memory from below. We have
\begin{align}
\label{eq:gain-lowerbound-as}
\mathcal{G}(g) \stackrel{a.s.}{=}&  \frac{N\bar{d}}{({\bar{d}}/{g}+r_\delta)N} \\
\label{eq:gain-lowerbound}
=\hspace{0.03in}&	\frac{1}{{1}/{g}+(1-{1}/{g}-\delta)} \\
=\hspace{0.03in}&	\frac{1}{1-\delta}
,\end{align}
where~\eqref{eq:gain-lowerbound-as} holds because the distance of the nodes from memory is $r_\delta$, asymptotically almost surely. Observe that as the number of memories becomes close to $N$, i.e., $\delta \rightarrow (1-\frac{1}{g})$, the gain $\mathcal{G} \rightarrow g$.
\end{IEEEproof}

\vspace{-.1in}\section{Network Compression in Internet-like Power-law Random Network Graphs}
\label{sec:wired-powerlaw}
In the previous section, using Erd\H{o}s-R\'enyi random graphs, we demonstrated that if the number of memory-enabled nodes $M$ increases like $N^{\frac{1}{g}}$, where $N$ is the total number of nodes, and $g$ is the memory-assisted compression gain, the network-wide benefits start to shine. Note that for any $g>1$, as $N \to \infty$ we have
\begin{equation}
\lim_{N \to \infty} \frac{N^{\frac{1}{g}}}{N} = 0.
\end{equation}
Therefore, even using an asymptotically vanishing number of memory-enabled nodes, we can obtain network-wide improvements. On the other hand, we expect even better network-wide scaling when the network has more structure than the Erd\H{o}s-R\'enyi random graphs.

The focus of the study of this section is to extend network compression to random power-law graph (RPLG) model.
The power-law graphs are particularly of interest because they are one of the useful mathematical abstractions of real-world networks, such as the Internet and social networks. In power-law graphs, the number of vertices whose degree is $x$, is proportional to $x^{-\beta}$, for some constant $\beta > 1$. For example, the Internet graphs have powers ranging from 2.1 to 2.45~\cite{Albert1999,Faloutsos1999,Broder2000,Internet1,Internet2}.
Accordingly, in the rest of this section we specifically direct our attention to power-law graphs with $ 2<\beta<3 $ (which include the models for the Internet graph), and provide results for memory deployment on such network graphs.
Our study entails first finding the optimal strategy for deploying the memory units and then investigating the effect of these memory units on the routing algorithms. The latter is important for the numerical evaluation of the network-wide gain as well. 

\subsection{Random Power-law Graph Model}

 A random power-law graph is an undirected, unweighted graph whose degree distribution approximates a power law with parameter $\beta$. Basically, $\beta$ is the growth rate of the degrees. To generate a random graph that has a power-law degree distribution, we consider the Fan-Lu model~\cite{chung-book}. In this model, the expected degree of every vertex is given. The Random Power-Law Graph (RPLG), with parameter $\beta$, is defined as follows:
\begin{definition}[Definition of $G(\beta)$]
Consider the sequence of the expected degrees $\boldsymbol {w} = \{w_1,w_2,\ldots,w_N\}$, and let $\rho = 1/\sum w_i$. For every two vertices $v_i$ and $v_j$, the edge $v_iv_j$ exists with probability $p_{ij}=w_iw_j\rho$, independent of other edges. If
\begin{equation}
\label{eq:weight}
w_i = ci^{-\frac{1}{\beta-1}} \text{ for } i_0\leq i \leq N+i_0,
\end{equation}
then graph $G(\beta)$ constructed with such an expected degree sequence is called an RPLG with parameter $\beta$. Here, the constant $c$ depends on the average expected degree $\bar{w}$, and $i_0$ depends on the maximum expected degree $\Delta$. That is,
\begin{equation}
\left\{
\begin{array}{lcl}
c	&=&	\frac{\beta-2}{\beta-1}\bar{w}N^{\frac{1}{\beta-1}},\\
i_0	&=&	N\left(\frac{\bar{w}(\beta-2)}{\Delta(\beta-1)}\right)^{\beta-1}.
\end{array}
\right.						
\end{equation}
\label{def:graph}
\end{definition}

With the definition above, it is not hard to show that the expected number of vertices of degree $x$ in $G(\beta)$ is $\approx x^{-\beta}$. In~\cite{chung-book}, authors showed that for a sufficiently large RPLG, if the expected average degree of $G(\beta)$ is greater than 1, then $G(\beta)$ has a unique giant component (whose size is linear in $N$), and all components other than the giant component have size at most $O(\log N)$, with high probability. Since we only consider connected networks, we will focus on the giant component of $G(\beta)$ and ignore all sublinear components. Further, by a slight abuse of the notation, by $G(\beta)$ we refer to its giant component.
Next, we briefly describe as to how the structure of RPLG provides insight about the efficient placement of memory-enabled nodes.

\vspace{-.07in}\subsection{Memory Deployment in Random Power-law Graphs}
Although memory deployment problem in a general graph is a hard one, the RPLG with parameter $2<\beta<3$ has a certain structure that leads us to finding a very good deployment strategy. The RPLG can be roughly described as a graph with a dense subgraph, referred to as the \emph{core}, while the rest of the graph (called periphery) is composed of tree-like structures attached to the core. Our approach to solve the memory deployment problem is to utilize this property and size the core of $G(\beta)$ and show that almost all the traffic in $G(\beta)$ passes through the core. We propose to equip all the nodes in the core with memory and hence almost all the traffic in $G(\beta)$ would benefit from the memories. The number of memories should be such that the network-wide gain is greater than 1 as $N\rightarrow \infty$. Showing that almost all the traffic goes through the core guarantees that $\mathcal{G}>1$ as shown in Lemma~\ref{lem:memory-on-path} below. This way, we find an upper bound on the number of memories that should be deployed in an RPLG in order to observe a network-wide gain of network compression. We will also verify that the number of memory units does not have to scale linearly with the size of the network to achieve this gain.

From Definition~\ref{def:graph}, we note that the nodes with higher expected degrees are more likely to connect to each other and also other nodes. Therefore, we expect more traffic to pass through these nodes. In our case, we are looking to size the core, i.e., find the number of high degree nodes such that almost all the traffic in the graph passes through them. Theorem~\ref{thm:core} below is our main result regarding the size of the core:
\begin{theorem}
\label{thm:core}
Let $G(\beta)$ be an RPLG. In order to achieve a non-vanishing network-wide gain $\mathcal{G}$, it is sufficient to deploy memories at nodes with expected degrees greater than $lw_{\min}$, where $l$ is obtained from
\begin{equation}
\label{eq:l-beta}
l^{3-\beta}-\frac{1}{\bar{w} \gamma}=0
,\end{equation}
and the constant $\gamma$ is equal to $(1- \frac{1}{\beta-1})^2 \frac{\beta-1}{3-\beta}$. The set of nodes with expected degree greater than $lw_{\min}$ is defined as core: $\mathcal{C}=\{u|w_u > lw_{\min}\}$.
\end{theorem}
Proof of the Theorem~\ref{thm:core} follows from the lemmas below.
\begin{lemma}
\label{lem:memory-on-path}
Let $d$ be the distance between the nodes $A$ and $B$. Let $\mu$ denote a memory unit fixed on the shortest path between $A$ and $B$, with distance $d'$ from $A$, i.e., the distance between $\mu$ and $B$ is $d-d'$. If the gain of memory-assisted compression is $g>1$, then $\mathcal{G}>1$.
\end{lemma}
\begin{IEEEproof}
If there was no memory on the path, we had one unit of flow from $A$ to $B$ and one unit of flow for $B$ to $A$. Therefore, $\mathcal{F}^0 = 2d$. When memory-assisted compression is performed, the flow going from $A$ to $B$ is reduced to $\frac{d'}{g}+(d-d')$. Similarly, the flow going from $B$ to $A$ is $\frac {d-d'}{g}+d'$. Therefore, $\mathcal{F} = \frac{d'}{g}+(d-d') + \frac {d-d'}{g}+d'$ and thus
\begin{equation}
\mathcal {G}(g) = \frac{2d}{\frac{d'}{g}+(d-d') + \frac {d-d'}{g}+d'}=\frac{2g}{g+1}.
\end{equation}
Now, considering that $g>1$, the claim follows.
\end{IEEEproof}

Our approach to find the core is to remove the highest degree nodes from the graph one at a time until the remaining induced subgraph does not form a giant component. In other words, as a result of removing the highest degree nodes, the graph decomposes to a set of \emph{disjoint} islands and hence, we conclude that the communication between those islands must have passed through the core.
Therefore, from Lemma~\ref{lem:memory-on-path}, we conclude that in RPLG, we will have a non-vanishing network-wide gain if we choose the core sufficiently big such that the induced periphery of $G(\beta)$ does not have a giant component. The following lemma provides a sufficient condition for not having a giant component in RPLG.
\begin{lemma}[\cite{chung-book}]
\label{lem:giant-component}
A random graph $G(\beta)$ with the expected degrees $\boldsymbol {w}$, almost surely has no giant components if
\begin{equation}
\label{eq:nogiantcomponent}
\frac{\sum_i w_i^2}{\sum_i w_i} < 1.
\end{equation}
\end{lemma}

\begin{lemma}
\label{lem:subgraph}
Consider a random graph $G$ with the sequence of the expected degrees $\boldsymbol {w}$. If $U$ is a subset of vertices of $G$, the induced subgraph of $U$ is a random graph with the sequence of the expected degrees $\boldsymbol {w'}$, where
\begin{equation}
w_i' = w_i\frac{\sum_{v\in U}w_v}{\sum_{v\in G}w_v}
.\end{equation}
\end{lemma}
\begin{IEEEproof}
The probability that an edge exists between two vertices of $U$ is equal to the edge connection probability in $G$. Consider a vertex $u$ in $U$. The expected degree of $u$ is
\begin{equation}
\rho\sum_{v\in U}w_uw_v = w_u \frac{\sum_{v\in U}w_v}{\sum_{v\in G}w_v}.
\end{equation}
\end{IEEEproof}

\begin{figure}[t]
\begin{center}
  \includegraphics[width=0.85\linewidth]{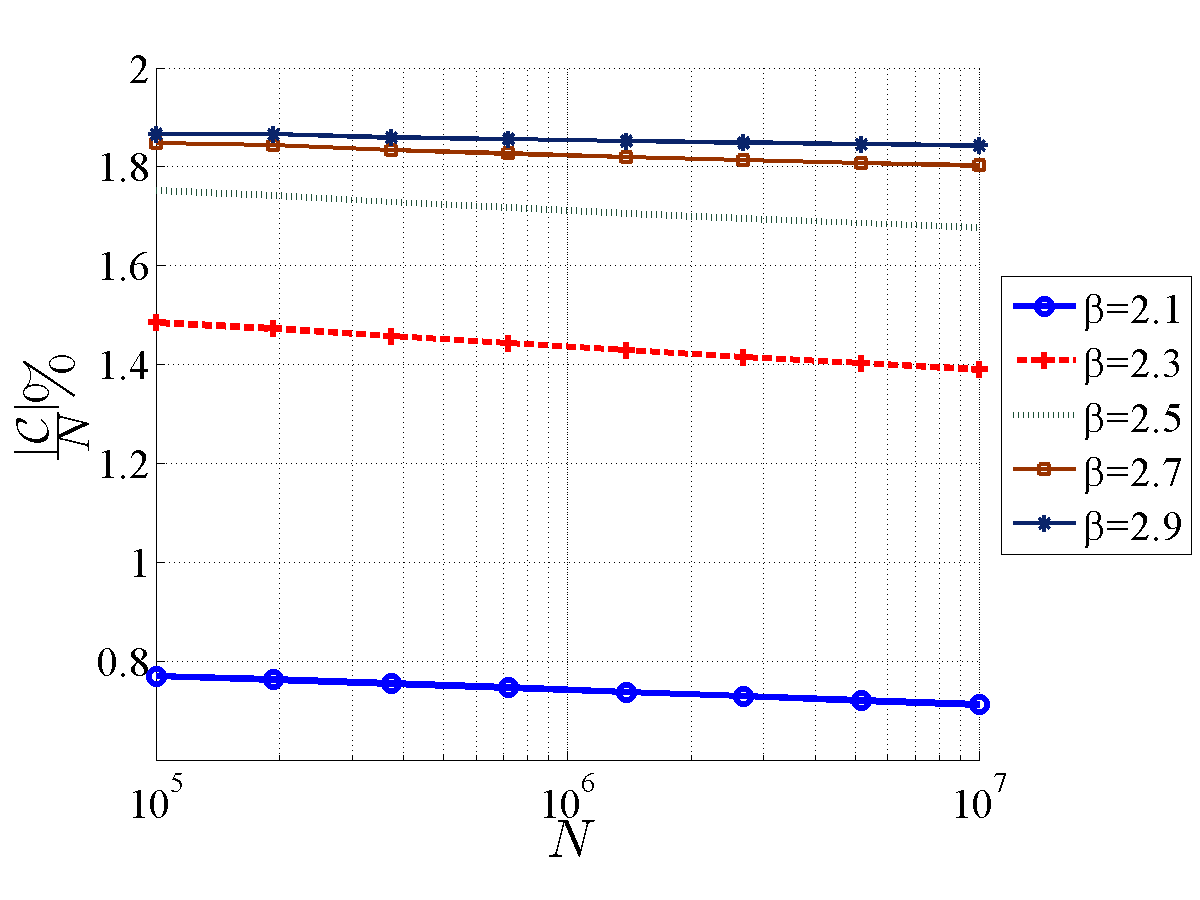}
\end{center}
\vspace{-.1in}
\caption{The scaling of the core size $\frac{|\mathcal{C}|}{N}\times 100$ versus $N$ for different $\beta$'s.}
\label{fig:U_l}
\vspace{-.1in}
\end{figure}

\begin{IEEEproof}[Proof of Theorem~\ref{thm:core}]
Consider a $G(\beta)$ with the set of lowest degree nodes $U_l$, all having expected degrees in the interval $(w_{\min},lw_{\min})$. According to Lemma~\ref{lem:giant-component}, to ensure that the induced subgraph $G_{U_l}$ does not have a giant component, we should have
\begin{equation}
\sum_{v\in U_l}{w_v'}^2/\sum_{v\in U_l}{w_v'} < 1
,
\end{equation}
where
\begin{equation}
w_v' = w_v\frac{\sum_{v\in U_l}w_v}{N\bar{w}}
\end{equation}
as in Lemma~\ref{lem:subgraph}. To find $w'$, we should first obtain $\sum_{v\in U_l}w_v$. According to~\cite{chung-book}, we have
\begin{equation}
\label{eq:volU_l}
\begin{array}{rl}
\sum_{v\in U_l} w_v 		\approx&\hspace{-0.08in} N \bar{w} (1-l^{2-\beta}),\\
\sum_{v\in U_l} {w_v}^2 	\approx&\hspace{-0.08in} N \bar{w}^2 (1- \frac{1}{\beta-1})^2 \frac{\beta-1}{3-\beta}l^{3-\beta}.
\end{array}
\end{equation}
From~\eqref{eq:volU_l}, we conclude that $w_v'= (1-l^{2-\beta})w_v,$ for all $v\in U_l$. Thus we have,
\begin{equation}
\label{eq:volU_l-2}
\sum_{v\in U_l} w'_v \approx N \bar{w} (1-l^{2-\beta})^2
.\end{equation}
Similarly,
\begin{equation}
\label{eq:vol2U_l-2}
\sum_{v\in U_l} {w'_v}^2 \approx N \bar{w}^2 \gamma l^{3-\beta}(1-l^{2-\beta})^2
.\end{equation}
Combining~\eqref{eq:volU_l-2} and~\eqref{eq:vol2U_l-2}, we obtain the relation in~\eqref{eq:l-beta} between $\beta$ and $l$. Having $l$, we can easily obtain the size of $U_l$ by finding the number of vertices with expected degree less than $lw_{\min}$ which is readily available from~\eqref{eq:weight}.
\end{IEEEproof}
Theorem~\ref{thm:core} provides the required information to find the size of the core and hence the number of memory units. As finding the closed-form solution for the size of the core is not straightforward, we use numerical analysis to characterize the number of required memory units using the results developed above.

\begin{figure}[t]
\begin{center}
  \includegraphics[width=0.9\linewidth]{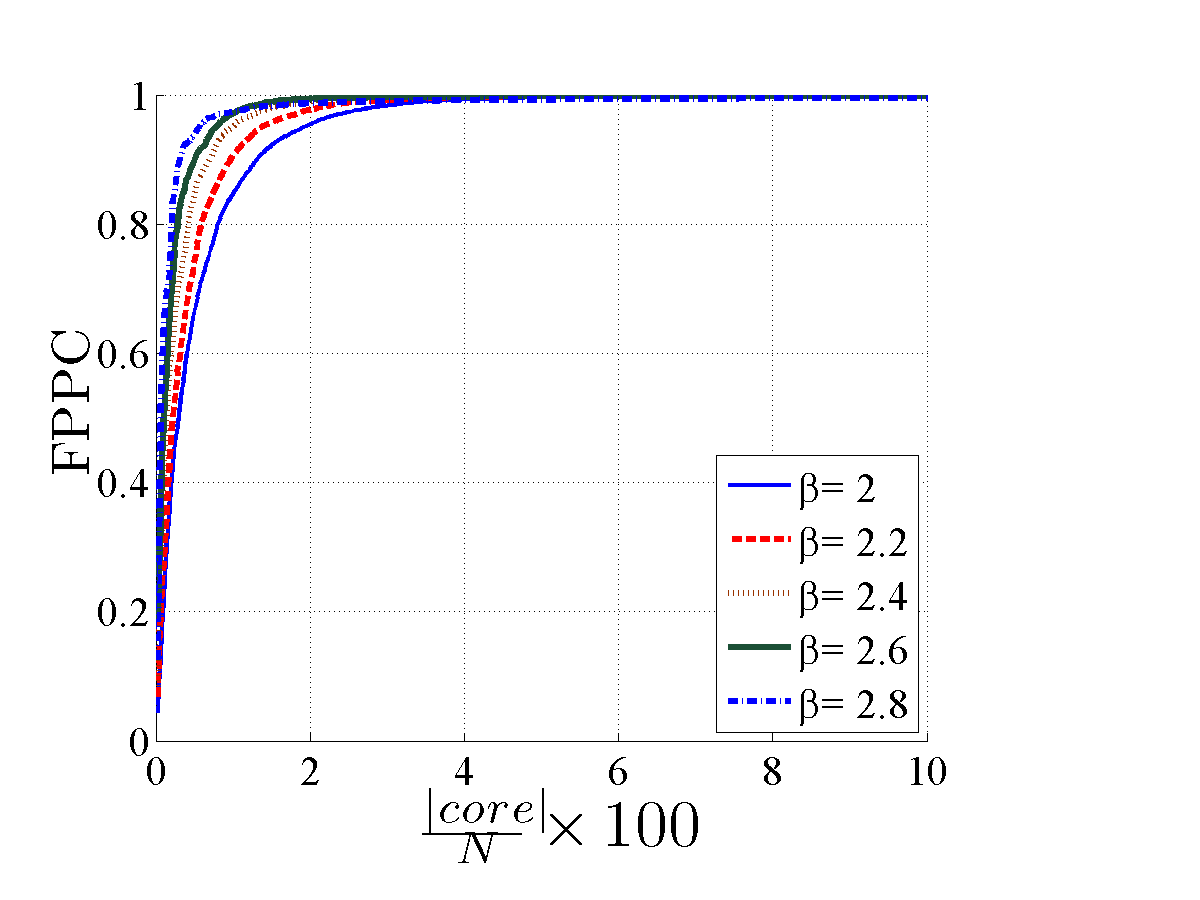}
\vspace{-.1in}
  \caption{The fraction of the paths passing through the core (FPPC) vs. the core size, for RPLG of size $N=5,000$.}
  \label{fig:path-through-core}
\end{center}
\vspace{-.1in}
\end{figure}

In Fig.~\ref{fig:U_l} the scaling of the core size versus $N$ is depicted for various $\beta$'s. As we see, the core size is a tiny fraction of the total number of nodes in the network and this fraction decreases as $N$ grows. This is a promising result as it suggests that by deploying very few memory units, we can reduce the total amount of traffic in a huge network.

\vspace{-.07in}\subsection{Simulation Results}
To validate our theoretical results, we have conducted different sets of experiments to characterize the network-wide gain of memory in RPLG. For experiments, we used DIGG RPLG generator~\cite{Brady2006}, with which we generated random power-law graph instances with number of vertices between 1,000 and 5,000, and $2<\beta<3$. The result are averaged over 5 instances of generated RPLGs. In our simulations, we report results for various core sizes (which is the number of memory-enabled nodes).

We first verify our assumption that a tiny fraction of the highest degree nodes observes most of the traffic in the network. Figure~\ref{fig:path-through-core} shows the Fraction of the Paths Passing through the Core (FPPC) for different core sizes and $\beta$'s. As we expected, more that $90\%$ of the shortest paths in the graph involve less than $2\%$ of the highest degree nodes to route the flow. Although our theoretical result in Theorem~\ref{thm:core} is asymptotic in $N$, Figure~\ref{fig:path-through-core} suggests that the asymptotic result holds for $N=1,000$ already. Therefore, we can place the memory units at the core and results can be extrapolated for large graphs with large number of nodes.

\begin{figure}[t]
\begin{center}
  \includegraphics[width=0.9\linewidth]{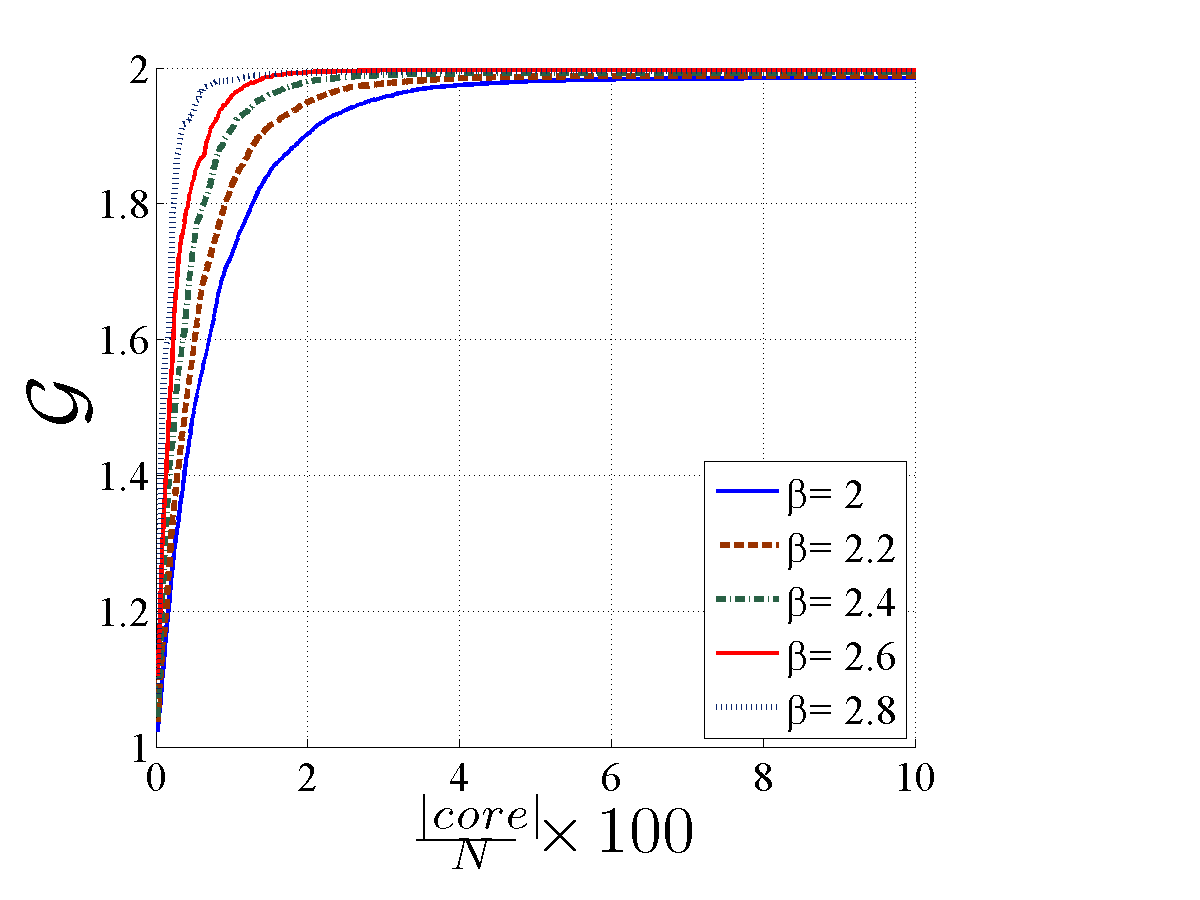}
\vspace{-.1in}
  \caption{Illustration of the network-wide gain when simple Dijkstra's routing algorithm is used for $g \to \infty$. Note that $\mathcal{G}$ is capped to two despite $g$ growing large.}
  \label{fig:G-suboptimal}
\end{center}
\vspace{-.2in}
\end{figure}

To validate the network compression gain, we considered two RPLGs with sizes $N=2,000$ and $N=4,000$. Assume that each memory node has observed a sequence of length $m=4$MB of previous communications in the network. The packets transmitted in the network are of size 1kB. This assumption is in accordance with the maximum transmission unit (MTU) of 1,500 bytes allowed by Ethernet at the network layer. From our results in Section~\ref{sec:ex-CNN}, we expect that a memory-assisted compression gain of $g\approx 2.5$ is achievable for real traffic traces. Hence, we use $g=3$ in our simulations. Please note that in the simulations of this section, the impact of the source statistics and compression is only considered through $g$.

\begin{figure*}[t]
\begin{center}
  \subfigure[$\frac{|core|}{N}=2.5\%$]{
  \includegraphics[width=.27\textwidth]{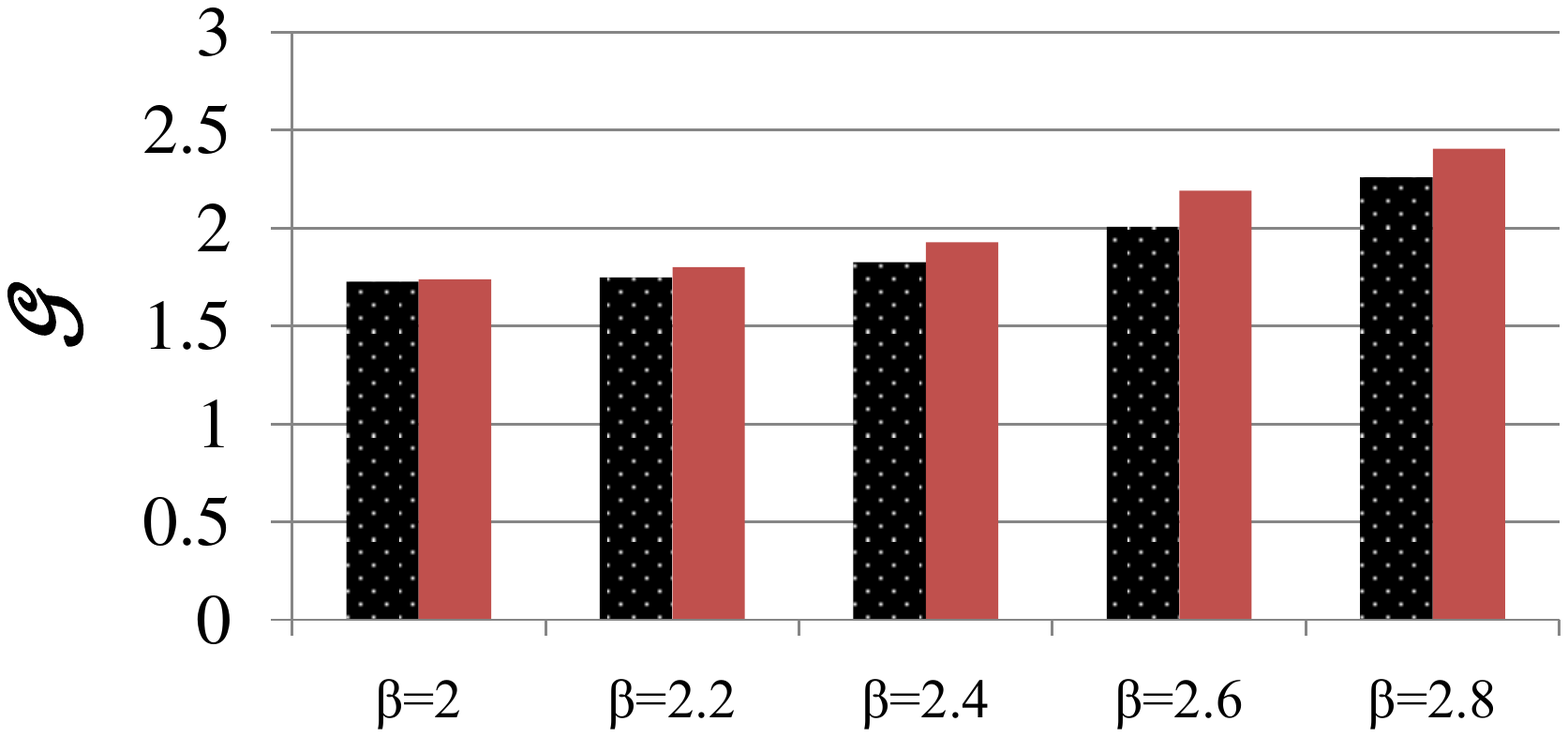}
  \label{fig:G-mem-2-5}
  }
  \subfigure[$\frac{|core|}{N}=5\%$]{
  \includegraphics[width=.27\textwidth]{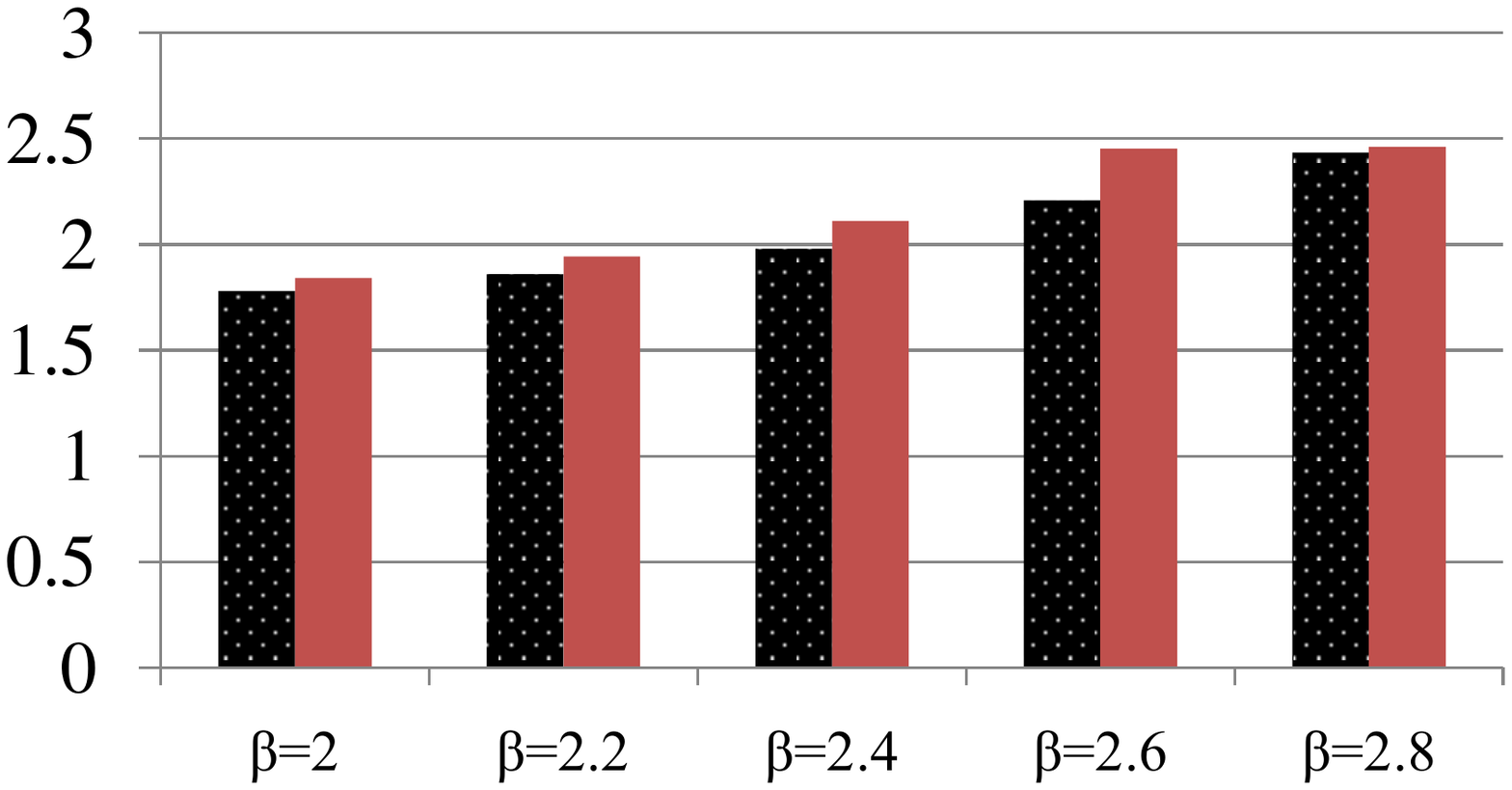}
  \label{fig:G-mem-5}
  }
  \subfigure[$\frac{|core|}{N}=10\%$]{
  \includegraphics[width=.27\textwidth]{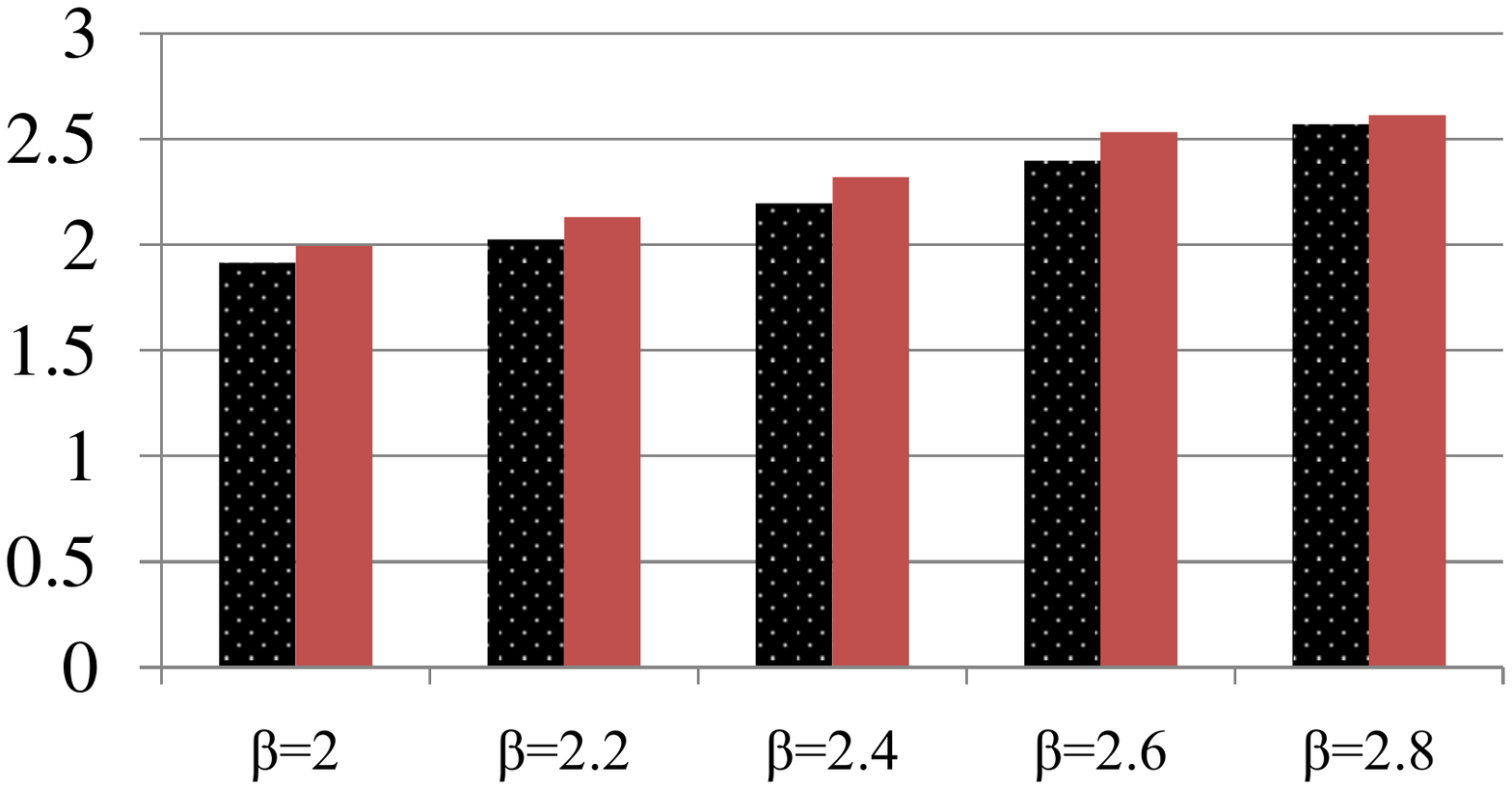}
  \label{fig:G-mem-10}
  }
  \subfigure{
  \includegraphics[width=.09\textwidth]{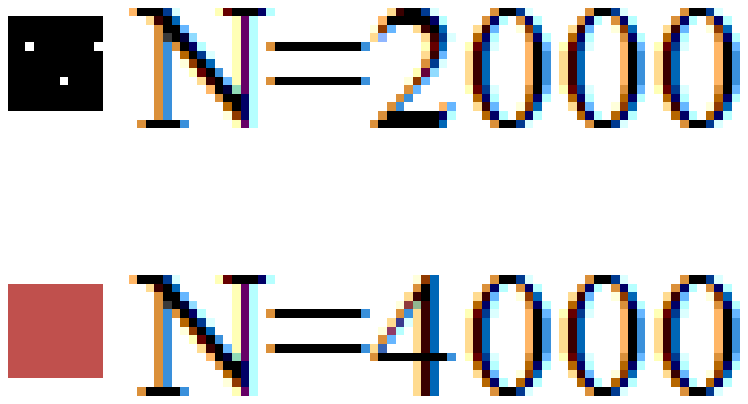}
  }
\end{center}
\vspace{-.1in}
\caption{Network-wide gain of memory assisted compression $\mathcal{G}$ for different core sizes and power-law parameter $\beta$, for $g=3$. }
\label{fig:network-wide-gains}
\vspace{-.1in}
\end{figure*}

To verify the results of routing with memory-enabled nodes in Section~\ref{sec:wired-routing}, we conducted the following experiment. If we do not use the modified Dijkstra's algorithm in the networks with memory (i.e., we do not optimize the routing algorithm to utilize the memories), as Lemma~\ref{lem:memory-on-path} suggests, the network-wide gain would be bounded by $\frac{2g}{g+1}$. Therefore, even for very large values of $g$, the network-wide gain would remain less than two (as shown in Fig.~\ref{fig:G-suboptimal}), which is not desirable. Figure~\ref{fig:network-wide-gains} describes our results for the achievable network-wide gain of memory-assisted compression. We measured the total flow without memory. We also obtain the optimal paths when we have memory units are deployed. We consider three cases in which the fraction pf nodes equipped with memory increases from $2.5\%$ of the nodes to $10\%$. All data has been averaged over the 5 graphs in each set.

The trendlines suggest that $\mathcal{G}$ increases as $\beta$ increases which is expected since the FPPC increases with $\beta$. In other words, more traffic between the nodes in periphery has to travel through the dense subgraph (core) as $\beta$ increases. Further, by increasing the number of memory units, the network-wide gain increases and approaches to the upper bound $g$. It is important to note that enabling only $2.5\%$ of the nodes in the network with memory-assisted compression capability, we can reduce the total traffic in the network by a factor of $ 2$
on top of flow compression without using memory, i.e., end-to-end compression. We emphasize that this memory-assisted compression (UcompM) feature does not have extra computation overhead for the source node (in comparison with the conventional end-to-end compression technique in Ucomp). Further, this feature only requires extra computation at the memory units when compared to Ucomp. The complexity of these operations scale linearly with the length of the data traffic. Hence, overall with some additional linear computational complexity on top of what could have been achieved using a mere end-to-end compression, network compression has the potential to reduce the traffic by a factor of $2$.

\vspace{-.1in}\section{Conclusion}
\label{sec:conclusion}
In this paper, we employed the concept of memory-assisted compression and introduced its implication in reducing the amount of traffic in networks.
The basic idea is to allow some intermediate nodes in the network to be capable of memorization and compression. The memory-enabled nodes observe the traffic traversing the network and form a model for the information source. Then, using the side information from this source model, a better universal compression of the flow is achieved on the network flow.
We investigated, from an information-theoretic point of view, the network flow compression by utilizing memory-enabled nodes in the network and solved the routing problem for networks with memory units. We also considered {\ER} random graphs and Internet-like power-law graphs to develop theoretical results on the number of memory-enabled nodes needed to obtain the network-wide benefits. 
Finally, our simulations demonstrated that by enabling memorization on less than $2.5\%$ of the nodes in an Internet-like random power law graph, we can expect almost all of the network-wide benefits of memorization providing two-fold gain over conventional end-to-end universal compression.

\section*{Acknowledgment}
The authors are thankful to Georgia Tech Networks and Mobile Computing (GNAN) Research Group for providing the traces of wireless network users used in the simulations of this paper. The authors also acknowledge very helpful discussions with Raghupathy Sivakumar about the benefits of combining de-duplication and memory-assisted compression. The authors are also thankful to the anonymous reviewers for their detailed comments that helped to improve the presentation of the paper.
\bibliographystyle{IEEEtran}
\bibliography{net-comp-all}
\end{document}